\documentclass[aps,twocolumn,superscriptaddress,longbibliography]{revtex4}

\pdfoutput=1
\usepackage[normalem]{ulem}

\usepackage[dvips,final]{graphicx}
\usepackage{amssymb}
\usepackage{amsmath}
\usepackage{amsfonts}
\usepackage{epsfig}
\usepackage{bm} % bold math
\usepackage[dvipsnames,usenames]{color}
\usepackage{comment}
\usepackage{float}
\usepackage[utf8]{inputenc}
\usepackage[colorlinks=true,linkcolor=blue,urlcolor=blue,citecolor=blue]{hyperref}
\usepackage[T1]{fontenc}
\usepackage{tikz-cd}
\usepackage{tikz}
\usepackage{booktabs}
\usepackage{mathbbol}

\usepackage{array}
\usepackage[caption=false,font=normalsize,labelfont=sf,textfont=sf]{subfig}
\usepackage{textcomp}
\usepackage{stfloats}
\usepackage{url}
\usepackage{verbatim}

\usepackage{xcolor}
\usepackage{xr}

\makeatletter
\newcommand*{\addFileDependency}[1]{% argument=file name and extension
  \typeout{(#1)}
  \@addtofilelist{#1}
  \IfFileExists{#1}{}{\typeout{No file #1.}}
}
\makeatother

\newcommand*{\myexternaldocument}[1]{%
    \externaldocument{#1}%
    \addFileDependency{#1.tex}%
    \addFileDependency{#1.aux}%
}

\myexternaldocument{S_information}

\begin{document}

\title[TQNNs and generalization]{Deep Neural Networks as the Semi-classical Limit of Topological Quantum Neural Networks:\\ 
The problem of generalisation}
\author{Antonino Marcian\`o}
\affiliation{Center for Field Theory and Particle Physics \& Department of Physics, Fudan University, Jingwan campus, Jingsan Rd, 200433 Shanghai, China and Laboratori Nazionali di Frascati INFN, Via Enrico Fermi, 54, 00044 Frascati (Rome), Italy, EU and INFN sezione Roma ``Tor Vergata'', 00133
Rome, Italy, EU}
\email{marciano@fudan.edu.cn}
\author{Emanuele Zappala}
\affiliation{Department of Mathematics and Statistics, Idaho State University\\
	Physical Science Complex |  921 S. 8th Ave., Stop 8085 | Pocatello, ID 83209}
\email{emanuelezappala@isu.edu}
\author{Tommaso Torda}
\affiliation{Department of Physics, University of Rome ``La Sapienza'', Piazzale Aldo Moro, 5, 00185 Rome, Italy, and the Laboratori Nazionali di Frascati INFN, Via Enrico Fermi, 54, 00044 Frascati (Rome), Italy}
\email{tommaso.torda@uniroma1.it}
\author{Matteo Lulli}
\affiliation{Department of Mechanics and Aerospace Engineering, Southern University of Science and Technology, 1088 Xueyuan Avenue, 518055 Shenzhen, China}
\email{lulli@sustech.edu.cn}
\author{Stefano Giagu}
\affiliation{Department of Physics, University of Rome ``La Sapienza'', Piazzale Aldo Moro, 5, 00185 Rome, Italy, and the Laboratori Nazionali di Frascati INFN, Via Enrico Fermi, 54, 00044 Frascati (Rome), Italy}
\email{stefano.giagu@uniroma1.it}
\author{Chris Fields}
\affiliation{Allen Discovery Center, Tufts University, 200 College Avenue, 02155 Medford, MA, US}
\email{fieldsres@gmail.com}
\author{Deen Chen}
\affiliation{College of Design and Innovation, Tongji University, 281 Fuxin Rd, 200092 Shanghai, China}
\email{deenchen@tongji.edu.cn}
\author{Filippo Fabrocini}
\affiliation{College of Design and Innovation, Tongji University, 281 Fuxin Rd, 200092 Shanghai, China and Institute for Computing Applications "Mario Picone", Italy National Research Council, Via dei Taurini, 19, 00185 Rome, Italy, EU}
\email{fabrocini@tongji.edu.cn}

%%%%%%%%%%%%

\begin{abstract}
%This is our abstract.
%%\vspace{0.5cm}
\noindent
Deep Neural Networks miss a principled model of their operation. A novel framework for supervised learning based on Topological Quantum Field Theory that looks particularly well suited for implementation on quantum processors has been recently explored. We propose using this framework to understand the problem of generalisation in Deep Neural Networks. More specifically, in this approach, Deep Neural Networks are viewed as the semi-classical limit of Topological Quantum Neural Networks. A framework of this kind explains the overfitting behavior of Deep Neural Networks during the training step and the corresponding generalisation capabilities. We explore the paradigmatic case of the perceptron, which we implement as the semiclassical limit of Topological Quantum Neural Networks. We apply a novel algorithm we developed, showing that it obtains similar results to standard neural networks, but without the need for training (optimisation).  
\end{abstract}

%\begin{IEEEkeywords}
%Topological quantum field theory, Generalization in neural networks, Topological quantum neural networks, Graph neural networks, Quantum amplitude %classifiers, Quantum perceptron.
%\end{IEEEkeywords}
% \maketitle

\maketitle

\section{Introduction}
\noindent
The problem of generalisation for deep neural networks (DNNs), i.e. neural networks with several hidden layers, is the problem of understanding how DNNs can successfully generalise. Test errors that are very close to their training errors can be achieved, even when they have sufficiently many model parameters to demonstrably ``memorise'' the training data \cite{zhang2017,belkin2018understand,poggio2017theory}.  In other words, the problem of generalisation is the problem of understanding why (at least some) overparametrised DNNs do not fail to generalise when they display an overfitting regime. Generalisation failure would be expected from the standard bias-variance trade-off; however, many DNNs evince a ``double-descent'' behavior, in which the generalisation performance increases, instead of decreasing, as the number of model parameters grows \cite{belkin2019reconciling}.  As shown by Zhang et al. \cite{zhang2017}, the methods of conventional generalisation theory, i.e. statistical learning theory, demonstrably fail to explain this behavior. The problem of generalisation has generated a rich literature \cite{34, 19, 20, 21, 22, 23, 24, 30, 25, 29, 8, 32, 33}, but remains unsolved.  As Kevin Hartnett \cite{hartnett_quanta} has written in Quanta Magazine:
\begin{quote} 
When we design a skyscraper, we expect it will perform to specification: that the tower will support so much weight and be able to withstand an earthquake of a certain strength. But with one of the most important technologies of the modern world, we are effectively building blind. We play with different designs, tinker with different setups, but until we take it out for a test run, we do not really know what it can do or where it will fail.
\end{quote}
~\\
\noindent
We do not really know, in fact, after one or even several test runs: it always remains the case that the ``next'' generalisation problem will reveal failure.  A principled approach for studying generalisation in DNNs that goes beyond conventional generalisation theory is, therefore, needed. 

Here we bring to completion our previous conjecture \cite{TQNN} that DNN architectures can be considered the semi-classical limit of a generalised quantum neural-network (QNN) architecture, the topological quantum neural network (TQNN). To this purpose, we elaborate on the theoretical framework we proposed in \cite{TQNN}, and provide empirical evidence to support our thesis, ultimately deriving a novel answer to the problem of generalisation.  TQNNs differ from conventional QNNs \cite{farhi2018classification, beer2020training} by allowing the number of ``layers'' and their connection-topology to vary arbitrarily, provided only that the input and output boundary conditions are preserved. The full generality of the TQNN framework as a representation of computational processes acting on classical data to yield classical outputs has been recently proven \cite{fgm:22}. We propose, in particular, that DNNs generalise {\em because} they are classical limits of TQNNs.  DNNs are, from this perspective, implementations of distributed, high-entropy, error-correcting codes that can be seen as classical limits of the quantum error-correcting codes (QECCs) \cite{knill:97} implemented by TQNNs \cite{fgm:23}. Increasing the model capacity of a DNN increases its generalisation ability because it increases the redundancy available in the code space. We demonstrate, using the perceptron architecture as a simplified example, that a TQNN can be used to directly compute the parameters needed to generalise from a training set, in the absence of any actual training of the network. This raises the possibility of replacing data-intensive training of DNNs with quantum computations, with a significant increase in efficiency and decrease in operational costs.

The plan of the paper is the following. In \S \ref{gen-theory}, we provide a formal definition of generalisation (\S \ref{formal-def}), and briefly review some expectations of conventional generalisation theory and how they have been observed to fail in DNNs (\S \ref{conventional-failure}). We then show, via a simple thought experiment, how generalisation can be expected to improve as parameters are added when a QECC is employed (\S \ref{thought-experiment}).  We provide a brief review of topological quantum field theory (TQFT), the theoretical framework underlying TQNNs, in \S \ref{tqft}, and summarise the TQNN construction in \S \ref{TQNN}. In \S \ref{sec:Gen_TQNN} we discuss the theoretical foundations of the correspondence between generalisation and the semi-classical limit of TQNNs. Specifically, in \S \ref{notion} we introduce the concept of generalisation as the semi-classical limit of TQNNs. In \S \ref{genqm} we discuss the paradigmatic case of the sum over quantum histories of a one-particle state, considered in analogy with a perceptron. We show that generalisation arises naturally from the path-integral formalism. In \S \ref{genTQNN} we extend this construction to any TQNN, by rephrasing the path-integral realising the sum over quantum histories for the perceptron, with generic spin-network states. In \S \ref{perce} we computationally showcase the theory of perceptron as a semi-classical limit of TQNN. In \S \ref{persp} we discuss possible perspectives to develop the current theoretical framework. Finally, in \S \ref{concl} we spell out our conclusions.

\section{Background and motivation} \label{gen-theory}

\subsection{The generalisation bound problem} \label{formal-def}

Informally, generalisation is the ability to identify members of some set (or class) $D$ after being exposed to the members of some proper subset $S \subset D$.  The generalisation ability of machine learning (ML) systems is tested by exposing them to new members of $D$ after training via backpropagation (or similar algorithm) on the members of $S$.  Whether they have identified new members of $D$ correctly is determined by human knowledge of $D$.  This knowledge is, however, not explicitly characterised: if we had an algorithmic specification of the members of $D$, we could encode that specification directly and ML would not be necessary.  We can make this dependence on non-algorithmic human knowledge explicit by defining generalisation as follows.

We can, without loss of generality, treat $D$ as a set of $N$-bit strings, i.e. $D = \{ 0,1 \}^N$.  Let $S \subset D$ be the training set.  Let $Y = \{ 0,1 \}^M$, with $M \leq N$, be the set of possible ``classes'' or ``answers'' to be obtained by acting on $D$.  In most cases of interest, $M \ll N$, e.g. classifying handwritten characters, distinguishing cats from dogs, or identifying individual people from photographs.

Now consider two agents $A$ and $B$, who implement functions $f_A, f_B: D \rightarrow Y$ and $g_A, g_B: S \rightarrow Y$ respectively, where we assume $g_A = f_A |_S$ and $g_B = f_B |_S$.  This latter assumption is true even if $A$ and/or $B$ is trained on a set $S^\prime \supset S$ that also contains noise or distractors, as in \cite{belkin2018understand}.  We can call $A$ the ``ground truth'' observer (e.g. a human) and $B$ the ``test'' observer (e.g. a ML system).  

The ``training error'' is $d(g_A, g_B)$, where $d$ is some metric on the relevant function space, and the ``test error'' or ``generalisation error'' is $d(f_A, f_B)$.  We can measure $d(g_A, g_B)$ because we can explicitly list every instance of $g_A$ and $g_B$ acting on $S$.  Both $f_A$ and $f_B$ are unknown, so we cannot compute $d(f_A, f_B)$.

We can now state the {\em generalisation bound problem} (GBP) as: given $d(g_A, g_B) \leq \epsilon$, determine an upper bound $\beta \geq \epsilon$ such that $d(f_A, f_B) \leq \beta$.  There is an {\em a priori} upper bound on $\beta$: $2^N$, corresponding to $A$ and $B$ disagreeing on every instance in $D$.  In this case, $\epsilon = 2^M$.  Our goal is to make $\beta \rightarrow \epsilon$ as $\epsilon \rightarrow 0$, which corresponds to both optimal learning and $S$ being optimally representative of $D$.

It is worth noting that the GBP cannot be solved in the general case.  The set $D$ may not be computable; we know, for example, that the generalisation from some set $S$ of programs that halt to the set $D$ of all programs that halt is not computable \cite{turing:37}.  There is, moreover, every reason to expect that human judges may disagree about edge cases for some classes of objects, and hence about membership in some candidate sets $D$, rendering test error in such cases observer-dependent.  It is also worth noting that humans are every bit as much black boxes as any ML system --- the ``explanation problem'' \cite{samek:21, taylor2021artificial} for humans is even harder than it is for ML systems, because experimenting on humans is more difficult than experimenting on machines.

\subsection{How conventional generalisation theory fails to explain DNN generalisation} \label{conventional-failure}

The goal of generalisation theory is to explain and justify why and how minimising the empirically measurable training error $d(g) = d(g_A, g_B)$ is a reasonable approach to minimising the test error $d(f) = d(f_A, f_B)$ by analysing the generalisation gap, i.e. $d(f) - d(g)$.  In practice, the generalisation gap is typically taken to be $f_B - g_B$, i.e. the ``gap'' is taken to characterise the performance of the ML system being tested.  This implicitly assumes that $g_A = f_A$, i.e. that the ``ground truth'' observer $A$ can identify all elements of $D$; we have seen above that this assumption can be incorrect.  With this assumption, the gap results entirely from the dependence of the trained predictor $g_S$ (here we drop the ``agent'' subscript for convenience) on the training set $S$. However, as noted above, $S$ might not be sufficiently representative of $D$ to direct the learner towards a good classifier for all of $D$.  In a ML context one is, moreover, more interested in the generalisation error than in the empirical training error; the main goal is evaluating the performance of the model on the unseen cases, i.e. on $D \setminus S$. If we look at the problem from this point of view, then the challenge comes from the mismatch between the optimisation task of minimising the empirical or training risk and the machine learning task of minimising the true or test risk.  Treating $D$ as a probability distribution over possible objects $x$, we can represent this mismatch as:
 
\begin{equation}
L_{D,f}(g_S)=P_{x\sim D} [g_S(x)\neq f(x)]= D[\{x: g_S(x)\neq f(x)\}], \nonumber
\end{equation}
where the error due to using $g_S$ is the probability of randomly drawing an example $x$, according to the distribution $D$, for which $g_S(x) \neq f (x)$, measured with respect to the probability distribution $D$ and the correct labeling function $f$. Here $P$ is the probability of a random variable, and $x\sim D$ represents sampling $x$ according to $D$.

The dependence of the predictor $g_S$ on the training data set $S$ has led to the definition of several examples of complexity measures providing bounds on the test error or on the sample-complexity. An instance of these measures are the VC-dimension or the Rademacher complexity. In particular, if the VC-dimension provides an upper bound on the test error, the Rademacher complexity measures the richness of a class of predictors with respect to a probability distribution. These results provide bounds on the test error that depend on the complexity or capacity of the class of functions $\mathcal{G}$ from which $g_S$ is drawn. In this sense, generalisation theory and capacity control are strictly related. Capacity control consists in using models that are rich enough to get good fits without using those which are so rich that they overfit.

The empirical success of DNNs, which are typically overparametrised models \cite{frankle2018lottery}, challenges the traditional complexity measures. Indeed, according for instance to the VC-dimension, the discrepancy between training error and generalisation error is bounded from above by a quantity that grows at least linearly as the number of adjustable parameters grows, but shrinks as the number of training examples increases. As a consequence, the traditional complexity measures tend to control the capacity by minimising the number of parameters \cite{vapnik1982necessary,blumer1989learnability,haussler2018decision}. Moreover, experimental results prove that DNNs that have been trained to interpolate the training data achieve a near-optimal test result even when the training data have been corrupted by a massive amount of noise \cite{belkin2018understand}. As Poggio et al. \cite{poggio2017theory} write, the main puzzle of DNN revolves around the absence of overfitting despite large overparametrisation and despite the large capacity demonstrated by zero training error on randomly labeled data. It is, in other words, the fact that DNNs can implement an operation $\mathcal{L}: (r,g_{S^\prime}) \mapsto f$, where $r$ is some initial, e.g. random, network state, that (at least approximately) correctly generalises to the desired $f$ from a function $g$ that is not the restriction of $f$ to the total training set $S^\prime$, but rather a noise function, namely a function of randomised data, on $S^\prime$. 

Historically, this puzzle was raised by a seminal paper of Zhang et al. \cite{zhang2017} mentioned earlier; see also \cite{neyshabur2014search}. This paper showed that successful deep model classes have sufficient capacity to memorise randomised data sets while having the ability to produce zero training error for particular natural datasets, e.g. CIFAR-10. The authors also empirically observed that explicit regularisation on the norm of weights seemed to be unnecessary to obtain small test errors, in contradiction to the ``traditional wisdom'' of generalisation theory. In the case of DNNs, the large capacity of the model looks sufficient to memorise the training data by brute force. This behavior conflicts with conventional generalisation theory, since learning by explicit memorisation of training examples should not imply generalisation capabilities. Generalising is traditionally understood as learning some underlying rule associated with the data generation process, i.e. learning some compact representation of the training function $g_S$, and therefore being able to extrapolate that rule from the training data to new unseen data. Moreover, as we have already mentioned, a result of this kind is a challenge to traditional complexity measures and, in general, to computational learning theory, since none of the existing bounds produces non-trivial results for interpolating solutions of the sort generated by such DNN models.

\begin{figure}[h!]%[DC]
\centering
\begin{center}
\includegraphics[width=8cm]{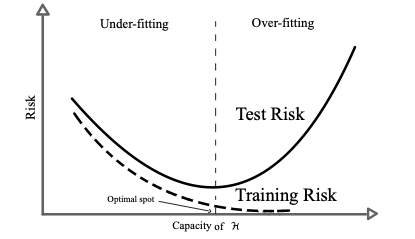}
%\end{center}
%\vspace{-0.6cm}  
\hspace{1cm}
%\begin{center}
\includegraphics[width=9cm]{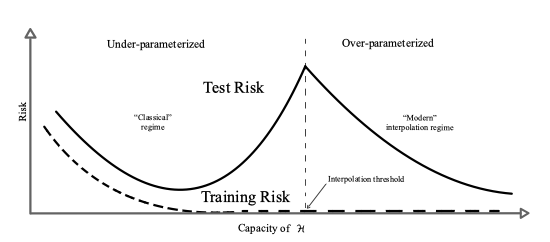}
\end{center}
\vspace{-0.6cm}
\caption{\footnotesize Traditional U-shaped risk curve, describing the trade-off between underfitting and overfitting, and double-descent risk curve, from \cite[Figure~1]{belkin2019reconciling}.}
\label{CRC}
\end{figure}

Several efforts have recently focused on achieving non-vacuous generalisation bounds for DNN models \cite{26, 24,arora2017generalization}. In the light of this situation, Belkin et al. \cite{belkin2019reconciling} propose to subsume the traditional U-shaped risk curve describing the trade-off between underfitting and overfitting with a double-descent risk curve Figure~\ref{CRC}. 
From the standard viewpoint, when the number of parameters $N$ is much smaller than the sample size $n$, i.e. $N<\!\!< n$, the traditional complexity measures assume that the training risk is close to the test risk. However, according to the double descent risk curve that the authors propose, by increasing progressively $N$ and thus by increasing the class capacity, the model will also increase the function classes until they are rich enough to achieve zero training error. As a consequence, near-perfect fit functions can be progressively constructed. These typically show a smaller norm and, thus, are simpler in the sense of the Occam’s razor. A view of this kind contradicts the traditional framework according to which, by increasing the class capacity, the predictors will perfectly interpolate the data. In this case, when an overfitting regime is achieved, a very high generalisation risk is achieved as well. Yet, as the authors state, increasing further the function class capacity beyond this point, leads to decreasing test risk, typically below the risk achieved when balancing underfitting and overfitting under the traditional approach.

% \begin{figure}[H]
% \centering
% \begin{center}
% \includegraphics[width=0.7\linewidth]{classical risk curve.png}
% \end{center}
% \vspace{-0.6cm}        
% \begin{center}
% \includegraphics[width=1.0\linewidth]{parameterized risk curve.png}
% \end{center}
% \vspace{-0.6cm}
% \caption{\footnotesize Risk curves in classical and modern regime.}
% \label{fig:strength}
% \end{figure}

This view matches recent results from Poggio et al. \cite{poggio2017theory} and Bartlett et al. \cite{bartlett2021deep} according to which overparametrisation leads to ``benign overfitting'' in which an accurate level of generalisation is achieved despite a near-perfect fit to training data. In particular, according to these authors, overparametrisation enables gradient techniques to impose regularisation implicitly while leading to accurate predictions despite their overfitting behaviour. Such behavior will prefer matrices showing a smaller norm, and hence a more distributed representation. This is a result that matches the thesis that the generalisation performance also depends on the size of the weights \cite{30}. Both of these assumptions guide our approach as well, even though they will be embedded into a theoretical framework totally different from classical complexity theory.

\subsection{A quantum approach to generalisation} \label{thought-experiment}

Quantum computing introduces a new and fundamentally nonclassical resource for the representation of data: quantum coherence, typically in the form of quantum entanglement \cite{nc:00}.  Just as classical computers employ additional bits --- from single-bit parity checks to full backup copies --- to detect and correct errors and hence enable robustness in the face of noise or other perturbations, quantum computers employ additional quantum bits (qbits) to implement QECCs. The extreme sensitivity of quantum states to environmental perturbations, which induce decoherence \cite{schloss:07} and hence ``erase'' data, requires that effective QECCs have large dimensionality, typically at least two qbits for every qbit to be protected~\cite{adh:15}.  Adding dimensionality, i.e. adding qbits to the code-space of the implemented QECC, to a quantum computer increases the range of perturbations against which the encoded data will be protected \cite{knill:97}.

The relevance of a high-dimensionality QECC to the generalisation ability of a QNN can be made obvious with a simple thought experiment.  Consider a two-player game in which $A$ and $B$ share a set $Y$ of classical tokens, and share a quantum channel $X$.  $A$ selects a token $y \in Y$ and encodes it into $X$ using some quantum operator $Q_A$, which is nondeterministic in the sense that $Q_A(y)$ depends on the immediately-previous state of $X$.  $B$ then uses an operator $Q_B$ to read a token $y^\prime$ from $X$.  $B$ sends $y^\prime$ to $A$, and $A$ sends back either `yes' or `no'.  $B$'s objective is to vary $Q_B$ so as to receive `yes' answers from $A$.  Hence asymptotically, $B$'s objective is to make $Q_B \rightarrow Q_A$.  

This game is an example of local operations, classical communication (LOCC) protocol \cite{chit:14}.  The local operations are the encoding and measurement steps that employ $Q_A$ and $Q_B$, respectively; the classical communication is the exchange of classical data ($y^\prime$) and `yes' or `no' answers.  The quantum channel subserving any successful LOCC protocol must implement a QECC to protect the quantum encoding of classical data \cite{fgm:23}.  In this game, $X$ implements a QECC, with $Q_A$ and $Q_B$ the encoding and decoding operators.  The dimension dim($X$) must, therefore, be much larger than dim($Y$); how much larger depends on what range of perturbations the code protects against.

Suppose $A$ and $B$ play this game for $n$ rounds, at the end of which $B$ is consistently getting `yes' from $A$.  What can be said about further rounds of the game? This is, clearly, the question of generalisation: the question of how similar $Q_B$ is to $Q_A$ after $n$ rounds of ``training''.  If the QECC implemented by $X$ protects against some set $B_\alpha$ of perturbations, and any encoding $x = Q_A(y)$ is such that $\exists B_\alpha ~{\rm such ~that}~ x = B_\alpha(x_i), x_i \in x_1 \dots x_n$, then $B$ will receive a `yes' after decoding $x$.  This corresponds to the $x_1 \dots x_n$ being representative of $Q_A(y)$ for arbitrary $y$ and arbitrary previous $|X \rangle$, given the set $B_\alpha$.  If we know the $B_\alpha$, we can find a lower bound on dim($X$).  Increasing dim($X$) increases the set of perturbations that $X$ can correct against.  Here it is clear that the generalisation problem is not solvable in the general case: a general solution of the generalisation problem would be a QECC that protects against a provably-complete set of perturbations mapping $S$ to $D$ for arbitrary $S$ and $D$.  If $D$ is unknown, no set of perturbations can be proved to be complete.

In this thought experiment, the training and test sets $S$ and $D$ are, effectively, measurements of the state in the ``middle'' of $X$, reflecting the fact that from a quantum-computing perspective, the ``world'' outside the communicating agents is a quantum information channel.  We can think of $A$ and $B$ --- a human and a ML system undergoing training --- as converging on a shared language --- a shared set of concepts.  The goal is for $B$'s concepts --- elements of $Y$ --- to refer to the same things as $A$'s concepts; they approach agreement asymptotically by playing the above game.  They each employ a quantum channel, the physical world, to generate a classical encoding, $D$.  They take turns showing each other an instance of $D$ and naming a concept.  This is how, for example, teaching a child to speak a language works.

In this quantum setting, there is no overfitting problem, because $X$ does not separate into components that individually encode particular inferences.  Instead, $X$ employs a high-dimensional entangled state to maximally distribute the representation of both the data and the implemented computation (in the above case, an Identity map).  A QECC is, in other words, as far from a classical look-up table as one can get; it effectively takes such a table and fully entangles it.  Hence for any channel that implements QECC, the test risk goes to zero as the dimensionality, and hence the range of perturbations the QECC protects against, increases.  

As mentioned earlier, DNNs are the classical limits of TQNNs, a generalisation of conventional quantum-gate based, layered QNNs \cite{TQNN}.  As quantum computers, TQNNs employ QECCs to achieve robustness \cite{fgm:23}.  Hence TQNNs, and DNNs as their classical limits, generalise better as their dimensionality increases.  Indeed minimising the norm of a DNN's matrix --- maximising the classical entropy of the representation --- is the classical limit of the use of entanglement to achieve a maximally distributed quantum encoding.

\section{Basics on Topological Quantum Field Theory} \label{tqft}

In this section we recollect some basic facts on TQFTs and give a general overview of the mathematical formalism used in this article. While TQFTs use the language of category theory, we will not delve into categorical approaches, but simply explain the meaning of certain categorical notions used in TQFT. We will also provide a physical perspective of TQFTs, which is the driving motivation of this work, as application of this framework to machine learning (in the semi-classical limit). 

\subsection{Mathematical foundation}\label{subsec:math_TQFT}

In general mathematical terms, using the Atiyah-Segal axioms \cite{atiyah1988topological,segal2001topological}, a TQFT is a functor from the category of cobordisms to the category of vector spaces. This definition, while very concise, is not particularly illuminating. We will therefore explain the most important points of this definition in a simplified manner.  

First, a cobordism is a manifold of dimension $n+1$ whose boundary is a union of manifolds of dimension $n$. As a simple example one can think of a cylinder. In fact, a cylinder is a manifold of dimension $2$, while a circle is a manifold of dimension $1$, and the boundary of a cylinder is a union of two circles. By category, in this context, it is meant that it is possible to perform operations between manifolds, such as glueing manifolds along their boundaries. One can therefore imagine, for example, to glue a cylinder to another cylinder along circles that constitute part of their boundaries. Therefore, we can compose manifolds by glueing along homeomorphic boundaries. This procedure is understood to be orientation-preserving. In addition, we have another type of operation, the ``tensor product'', which consists of taking the disjoint union of manifolds. 

The category of vector spaces in the context of TQFT simply refers to the class of vector spaces over a given field, e.g. $\mathbb C$, and the linear maps between them. Composition is the usual composition of maps. The tensor product, which is the second operation defined for cobordisms, is just the regular tensor product of vector spaces and linear maps. 

With these definitions in place, we are in the position to state what a functor between cobordisms and vector spaces is, and therefore to give a more descriptive definition of TQFT. Such a functor is an assignment of $n$-dimensional manifolds to vector spaces, and cobordisms between manifolds to linear maps between the corresponding vector spaces. One can think of a functor as a translation of the terminology from a category (the one of cobordisms in our case) to another category (the one of vector spaces in our case). The correspondence should satisfy certain coherence axioms, in the sense that compositions are sent to compositions, and tensor products are sent to tensor products.

If we denote by $Cob$ the category of cobordisms, and by $Vec_{\mathbb k}$ the category of vector spaces over a field $\mathbb k$, then a TQFT is a functor $\mathcal F: Cob \longrightarrow Vec_{\mathbb k}$, meaning that $\mathcal F(M)$ is a vector space for each $n$-dimensional manifold $M$, and any cobordism $\Sigma$ with boundary $M_1\sqcup M_2$ corresponds to a linear map $\mathcal F(\Sigma) : \mathcal F(M_1) \longrightarrow \mathcal F(M_2)$. The coherence axioms translate into simple equations such as
\begin{eqnarray}
    \mathcal F(\Sigma_1\circ_M \Sigma_2) = \mathcal F(\Sigma_1) \circ \mathcal F(\Sigma_2),
\end{eqnarray}
where $\Sigma_1\circ_M \Sigma_2$ is a cobordism (i.e. the $(n+1)$-dimensional manifold) obtained by glueing $\Sigma_1$ and $\Sigma_2$ along some boundary $n$-dimensional manifolds $M_1$ and $M_2$ (homeomorphic to $M$), and composition on the right is simply composition of maps. The fact that $\mathcal F$ respects tensor products becomes
\begin{eqnarray}
    \mathcal F(M_1\sqcup M_2) = \mathcal F(M_1)\otimes \mathcal F(M_2),\\
    \mathcal F(\Sigma_1\sqcup \Sigma_2) = \mathcal F(\Sigma_1) \otimes \mathcal F(\Sigma_2),
\end{eqnarray}
where the disjoint union of manifolds represents the tensor product of cobordisms, and tensor products on the right are assumed to be taken over the ground field $\mathbb k$. Figure~\ref{fig:reverse_pants} shows a simple example, where a $2$-dimensional manifold with a boundary consisting of three circles, the reversed pair of pants, corresponds to a linear map from a tensor product of two vector spaces to a vector space. 

\begin{figure}[h!]
    \centering
        \begin{tikzpicture}[scale = 0.35]
            \draw (-4,4) ellipse (2cm and 1cm);
            \draw (4,4) ellipse (2cm and 1cm);
            \draw (0,-4) ellipse (2cm and 1cm);
            \draw[rounded corners] (-6,4) ..controls(-5.5,2) and (-2,1).. (-2,-4);
            \draw[rounded corners] (6,4) ..controls(5.5,2) and (2,1).. (2,-4);
            \draw[rounded corners] (-2,4) ..controls(-1,1) and (1,1).. (2,4);
            \node (a)  at (8,0) {$\mapsto{\mathcal F}$};
            \node (a)  at (15,0) {$\begin{matrix} V & \otimes & V \\ & & \\ & \downarrow & \\ & &\\ & V & \end{matrix}$};
        \end{tikzpicture}
    \caption{A $2$-dimensional manifold (reversed pair of pants) with $1$-dimensional boundaries is mapped by a TQFT, $\mathcal F$, to its corresponding linear map. On top, the boundary consists of two circles, and therefore the domain of the corresponding linear map is a tensor product of vector spaces, while the bottom consists of a single circle, and the linear map has a single vector space as target.}
    \label{fig:reverse_pants}
\end{figure}

The interest of TQFTs in quantum topology and quantum algebra comes from the fact that manifolds can be decomposed (handle decomposition) into simpler components. The correspondence of these latter ones to algebraic counterparts through a functor $\mathcal F$ allows us to translate topological objects into algebraic ones. Moreover, when we consider a closed manifold, a TQFT $\mathcal F$ will make this correspond to a linear map $\mathbb k \longrightarrow \mathbb k$, since the boundaries will be empty. Such a linear map would simply be an element of the ground field, which is a ``numerical'' topological invariant of the manifold.  

In practice, to construct a TQFT one constructs algebraic objects that are invariant with respect to some way of presenting topological structure, such as simplicial decomposition through Pachner moves \cite{pachner1991pl}, link surgery presentations \cite{lickorish1962representation,wallace1960modifications}, handle decomposition \cite{kock2004frobenius}. The numerical topological invariant of a manifold obtained through a TQFT is usually said to be a partition function, in analogy with statistical mechanics, as the algebraic invariants are obtained through a summation over all possible compatible states of the topological structure in a way that while each single element changes based on the chosen combinatorial representation, the total sum is only dependent on the topological homeomorphism class.

\subsection{Physical interpretation}

The physical interest in TQFT arises in terms of low-energy approximations to physical theories. Arguably, or ``undoubtedly'' according to Atiyah in \cite{atiyah1988topological}, TQFTs have been motivated by Witten's study of supersymmetry in geometric terms \cite{witten1982supersymmetry}. In this context, the main objective of study in quantum field theories are moduli spaces (e.g. the space of connections) on some topological space. TQFTs, then, represent topological information that is robust upon taking the classical limit, and that therefore can be studied even when the ``complete'' quantum field theory is not known, and can motivate the formal construction of the latter.

With reference to the axiomatic definition given in Section~\ref{subsec:math_TQFT}, the correspondence between boundary and ground vector space represents the correspondence between physical space and Hilbert space of the quantum theory. A cobordism $(n+1)$-dimensional manifold between boundary $n$-dimensional manifolds indicates the addition of time to the physical space. For instance, we can think of a $3D$ theory with $4D$ spacetime. A cylinder $M\times I$ with two boundaries $M_i = M\times \{i\}$, $i = 1,2$ is a topologically trivial spacetime whose corresponding map between the Hilbert spaces $\mathcal F(M_1)$ and $\mathcal F(M_2)$ will be trivial. This situation corresponds to an evolution operator with trivial Hamiltonian, and therefore no dynamics of propagation \cite{atiyah1988topological}. However, even between homeomorphic $M_1$ and $M_2$, we can have cobordisms that are not topologically trivial --- e.g. a genus $1$ surface between two circles. This situation gives rise to nontrivial dynamics of propagation. 

The notion of numerical topological invariant of Section~\ref{subsec:math_TQFT} has an interesting interpretation in terms of physics as well: this is a probability amplitude for a vacuum to vacuum transition. 

In terms of physical interpretation, TQFTs are usually required to satisfy stricter axioms, such as the involutory axioms, relating the opposite orientation of a manifold to the dual space, and the Hermitian axiom referring to the adjoints. We refer the reader to \cite{atiyah1988topological,witten1982supersymmetry} for background on the motivation of such restrictions.

\section{TQNN in TQFT}\label{TQNN}
\noindent 
Before tackling the issue of generalisation, elaborating on the results attained by Zhang et al.,
\cite{zhang2017,zhang},
we summarise in this section the novel strategy rooted in the framework provided by TQFT \cite{TQNN}. This is an effective quantum theoretic approach that offers the pathway to addressing the problem of generalisation. We will show that the origin of this problem can be related to the topological encoding within the network structure, achieved through path selection and parameter optimisation,  
which coincide, in the semi-classical limit, with the training of classical neural networks. 
In addition, we argue that TQNNs naturally implement, through the selection of topological features, an optimisation over the architecture of the corresponding classical neural networks in the semi-classical limit. This is in stark contrast with the standard/classical case, where the architecture of the network is fixed beforehand, and the weights are learned during training. Consequently, optimal architecture selection and generalisation become, in the semi-classical limit, different interpretations of the same underlying TQNN procedure.
 
Within the framework of TQNN introduced in \cite{TQNN}, both the elements of the training and test samples are associated to quantum states --- the boundary states of the underlying TQFT --- that are colored with irreducible representations of Lie groups. %{\bf (we refer to these as metric data)}. 
In practice, these boundary states are represented by spin-networks that are dual to triangulations of a manifold. They encode the discretisation of space-time in that they combinatorially represent the continuous underlying manifold. %{\bf (we refer to the related combinatorial information as topological data)}. 
In this context, the $G$-bundle structure of the manifold is encoded in terms of group elements on the edges of the spin-network. 
Quantum states are represented by cylindrical functionals of the boundary group elements that are associated to the input data, and are supported on boundary graphs (1-complexes). Output boundary states represent instead the TQNN's ability to react to solicitations provided by the input (training/test samples). The functorial evolution from input to output boundary states is captured by 2-complexes, which realise the sum over histories proper of path integrals in quantum mechanics. The physical scalar product between boundary states is then used to produce a numerical output that later determines the training and test errors.

On the one hand, the ``input to output'' functorial evolution realises a sum over all the geometries of the system, similarly to the Misner-Hawking integral (see \cite{rovelli2011simple}). Certain geometries resonate giving rise to dominating terms in the output of the TQNN. In other words, the structure of the data induces TQNNs to select certain geometries. On the other hand, the learning procedure here consists of an optimisation on the parameters of the heat kernel corresponding to coherent spin-network states (see e.g. \cite{bianchi2010coherent}). Therefore, the resonances are determined by the integral over the histories learned during training. This duplicity reflects the double role of learning in TQNNs, where both the architecture and the weights are learned, as mentioned above. We point out, additionally, that in the present article we consider the semi-classical limit of TQNNs where no optimisation is performed. Our procedure is based on a direct computation that determines, with no optimisation or learning involved, the saddle point configuration of the Feynman path integral. Therefore, our approach in this article is optimisation-free. However, we give in the rest of this section a general perspective for TQNNs (with no semi-classical limit) where optimisation can be eventually performed, in order to better illustrate the theoretical framework from which this article has emerged.

Our approach follows the very same axioms of quantum field theories \cite{TQNN}, and hinges on the following steps:

\begin{enumerate}
\item 
To data we associate (possibly a superposition of) spin-networks through some encoding procedure (see \cite{TQNN}). The spin-networks are determined by the irreducible representations that label their edges and live in the boundary Hilbert space of the TQFT theory.

\item
The generic boundary states are characterized by two classes of parameters, which we dub as {\it topological} and {\it metric} parameters. The former ones are determined by the topology of the graph supporting the spin-networks and the topology of the manifolds, whose simplicial decompositions are dual to the spin-networks. These are quantum invariants obtained through state sums of admissible states. The metric parameters relate to the differential structure of the manifold and appear in the spin-network formalism from the fact that spin-network edges support the connection of the underlying manifold. These parameters are captured by the spin of the representation itself.

\item
Information provided by the training samples, together with the analogical definition of training and test error, in terms of the internal product of boundary quantum states, allows to fix the functorial structure of the bulk of the TQNN, namely the topological structure of the TQNN 2-complex. This procedure determines the topological parameters, since it gives the quantum groups in terms of which the topological invariants are expressed. We also observe that the weights of the boundary coherent states are learned during the training process.  

\item 
The topological parameters are enough to capture the pattern underlying the training set; whenever the ``information'' about the topology specified by the training data is not sufficient and/or noise exceeds the possibility to individuate topological structures, an interpolating pattern which does not select specific topological features is then deployed.

\item
The metric parameters are individuated by the Gaussian weights associated to the coherent group elements assigned to the TQNN states. In the semi-classical limit these weights correspond to the matrix weights of DNNs. For instance, in \cite{TQNN} we recover the perceptron as the semi-classical limit of a specific TQNN.

\item 
The measure of the Hilbert spaces associated to the links, that are deemed to be equivalent to the training sample set, characterises the minimal amount of information flow required to achieve pattern identification: if not enough information is provided by the group elements, which is determined through the association of spin-network states to input data, not enough information is provided to reconstruct the topological invariants.

\item 
The ``richness'' or ``energy'' of the irreducible representation sets allows to ``switch on'' the links, and thus the nodes and the topological linking and knotting invariants, only for non-trivial (non-zero) values of the spin representations. Thus, the analogue of the minimal length in theories of quantum gravity, or Planck length, is the minimal spin, or information bit, in the framework of \cite{TQNN}.
\end{enumerate}

Relying on this framework, we propose in the next section that generalisation, within the context of standard/classical DNNs, emerges from the optimisation in the semi-classical limit of the TQFT path-integral (state sum), deployed to estimate the classifier, as a direct consequence of the classical path selection at the saddle point. Moreover, topological features, including topological invariants (state sums) of the underlying TQFT and graph connectivity of spin-network boundary states, are essential elements to achieve generalisation. In fact, as argued in more detail in Section~\ref{sec:Gen_TQNN}, we have that learning corresponds to singling out preferred topologies in the form of resonances in the transition amplitudes. These are dominant terms in the partition function induced by the path-integral of the TQFT.

Change of the graphs' topology is achieved at the (infinite number of quantum) hidden layers by vertices structures implementing the TQNN evolution. Through the topological features of the 1- and 2- complexes, the TQNN can capture the topological invariants from the training sets. On the other hand, parameter optimisation in the state sums corresponds to the individuation of the semi-classical limit of the quantum theory: the ``classical'' path is individuated by the extremisation of the action of the theory in the sum over histories; fluctuations around that path still enable to achieve generalisation, as a minimisation of the training error. 

\begin{figure}[h!]%[DC]
\centering
%\begin{center}
% \includegraphics[width=9 cm
%0.6
%\linewidth
% ]{DoubleComplexity-v3-2.pdf}
\includegraphics[width=9 cm
]{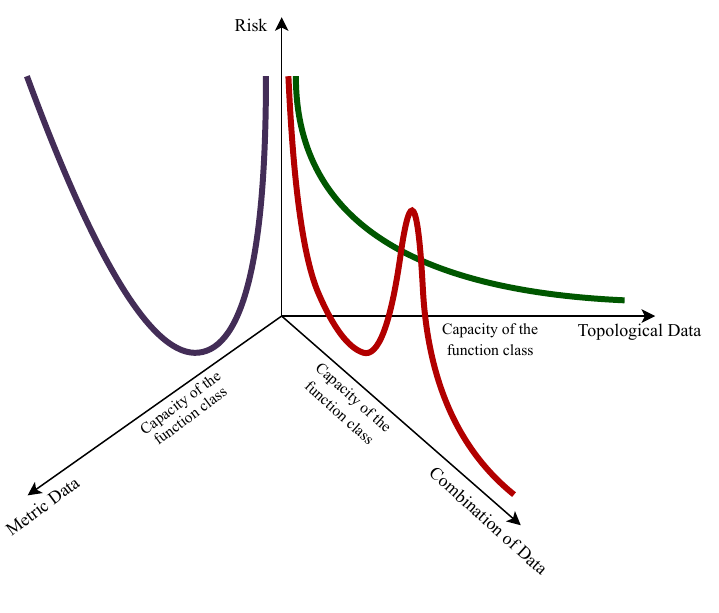}
% ]{Uno.png}
%\end{center}
\caption{\footnotesize The empirical double-descent curve (red) is reproduced when both topological and metric complexity are taken into account.}
\label{DC}
\end{figure}

Within this novel picture, topological and metric data captured by the TQNNs' structure increase the class capacity and thus ultimately the function classes of the quantum ML models under scrutiny. Indeed, refining the triangulation/tessellation of geometric manifolds underlying the description of data ensembles results into an increase --- either in the simplicial skeletons extracted by manifolds triangulation/tessellation or in their dual 1-complexes --- of the topological connectivity of the graphs on which TQNNs are supported. While increasing the ensemble of metric data saturates the classical risk curve to provide a standard U-shaped form, the complexity of the TQNNs graph structure that encodes the  topological data provide an asymptotic improvement of generalisation. For TQNNs graph structures that are not sufficiently rich, the model, only characterised by the flow of metric data along the links of the 1-complexes, undergoes overfitting as the ensemble of metric informations increases. Nonetheless, increasing the graph complexity through enlarging the topological data ensembles enables metric data to increase function capacity. This suggests that the one-dimensional curves that appear in Figure~\ref{CRC} should be replaced by a curve that takes into account two different independent classes for the parameters characterising the models under scrutiny, to be represented respectively on the axes for the metric and the topological complexity, as depicted in Figure~\ref{DC}.\\

The double descent is obtained as the product of the metric and the topological information. Figure~\ref{DC} then suggests that the (spin-network) quantum states that encode both metric and topological information result from the product of states that can be visualized as occupying different orthogonal axes. Then their related amplitudes and statistical distributions will turn out to be composed multiplicatively. The standard U-shaped function is the result of interpolating among the two cases of underfitting (for low complexity) to overfitting (for high complexity). The fast-decreasing behaviour in the topological information  captures qualitatively the decrease of the accuracy error in the inverse of the volume of the total Hilbert space of the TQNN framework. This latter scales like the entropy measure of the spin-network quantum states' Hilbert space, factorially in the colors of the spin-network quantum states, and exponentially in their size. The double descent curve emerges in this scenario as due to hidden (topological in nature) properties of the dataset that are not taken into consideration in the classical case, when a U-shaped curve is obtained.

\section{Generalisation in the framework of TQNN}\label{sec:Gen_TQNN} 
\noindent 
Our perspective relies on the conjecture that generalisation happens as the analogue of the macroscopic manifestation of quantum mechanical effects. Specifically, generalisation can be addressed as the manifestation of topological (quantum) encoding achieved in TQNN and classical path-selection.

The texture of the webs of vertices and edges, which is determined by the data and captures the topology of the (triangulated) manifold as its dual simplicial complex, concretely implements a higher level topological pattern \cite{TQNN}, unveiling which path coincides with the procedure of learning.
Generalisation is consequently achieved in this framework as the selection of topological invariants, the most adequate to the achievement of a specific task. These topological invariants are captured in the Topological Quantum Physics (TQP) framework by the linking and knotting quantum numbers assigned to the graphs and to the 2-complexes of TQNN. 
%%
%%
%The network of connections between reference features or landmarks in a face of other image is an example of such an invariant~\textcolor{red}{[this last sentence does not seem clear]}; in principle, a TQNN represents all possible such invariants, up to its encoding capacity.

\subsection{The notion of generalisation in DNNs as semi-classical limit of TQNNs} \label{notion}
\noindent 
Let us now consider in detail the issue of generalisation in TQNNs, and consequently attempt to answer the problem raised in \cite{zhang} for DNN. 
When a coherent states representation of the Hilbert space is chosen, this corresponds to the selection of a family of states, in the training sample dataset, which is characterized by a random assignment of the irreducible group representation labels. The representation labels are in turn peaked around the labels of the coherent states that capture the mean values of the observable quantities and are determined through the quantum annealing procedure. This represents therefore a natural definition of labels’ randomisation in the training set. Label randomisation can be compared to randomly defining the elements of the Hilbert space, as this corrupts the correspondence between underlying data and correct label.

A classical DNN has only to learn the function $f^{\prime}$ in the learning algorithm $\mathcal{L}:(r,f')\rightarrow f$, which specifies example input-output pairs; it has no access to the ``intrinsic'' structure of the training examples.  TQNNs, however, are sensitive to such intrinsic structure in the form of topological features.  The structure of TQNNs naturally encode topological charges through the functorial quantum dynamics ensured by the 2-complexes, which create either vertices and then functions of $j$-representation and intertwiner quantum numbers, i.e. amplitudes, or other topological charges encoded in the knotting and linking of the edges in the bulk of the 2-complex. 

For clarity, let us consider the case of the TQNN analogue of the classical architecture of {\it perceptron}. As shown in \cite{TQNN} Section~5, the classical perceptron is recovered as the semi-classical limit of a TQNN. In fact, considering the composability properties of TQNN, inherited from the functoriality of TQFTs (the Atiyah axioms), it follows that this procedure describes also the {\it multilayer perceptron} (MLP) in the semi-classical limit. In particular, we argue that the semiclassical limit, which singles out the classical trajectories from the infinite amount of quantum trajectories, corresponds to the minimisation of the weight vector. For perceptrons, the components of the weight vector are determined by the minimisation of the loss functional or the learning task at hand. In the framework of TQNN, the weight vector is substituted by the set encoding the values of the action at the nodes and links of the boundary states --- see \cite{TQNN} Section~5 --- and at the infinitely many quantum hidden layers interpolating among the boundary states. The minimisation process that constitutes learning, therefore, in this case consists in the minimisation of a Feynman path-integral (associated to the TQFT), and it therefore amounts to selecting the dominant contributors to the infinitely many paths.

When the action is minimised, the weight amplitudes of the path integral are maximised. On the other hand, minimisation of the action ensures selection of the classical paths. In turn, tiny fluctuations around these minima correspond to the semi-classical paths nearest to the classical ones.

\begin{figure}[h!]
\centering
\begin{center}
\begin{tikzpicture}
\draw (2,5) ..controls (-1,4) and (4,2).. (0,0);
\draw (2,5) ..controls (3,3) and (1,2).. (0,0);
\draw[fill=black] (2,5) circle (2pt);
\draw[fill=black] (0,0) circle (2pt);
\draw[->,ultra thick] (-1,0)--(5,0) node[right]{$x$};
\draw[->,ultra thick] (-1,0)--(0-1,6) node[above]{$t$};
\node (a) at (2.65,5) {$(x_1,t_1)$};
\node (a) at (0,-.5) {$(x_0,t_0)$};
\draw[dashed] (-.5,4) -- (4.5,4) node[right]{$t$};
\draw[dashed] (-.5,4-1) -- (4.5,4-1) node[right]{$t'$};
\draw[dashed] (-.5,4-2) -- (4.5,4-2) node[right]{$t''$};
\draw[dashed] (-.5,4-3) -- (4.5,4-3) node[right]{$t'''$};
\end{tikzpicture}
\end{center}
\vspace{-0.6cm}        
\caption{\footnotesize  Possible paths in $xt$-plane. }
\label{fig:multiplepaths}
\end{figure}

The case of DNNs like MLPs consists of composing (via functoriality of the given TQFT) single units as in the case of the perceptron. The paradigm described above is substantially unchanged.

\subsection{Generalisation for the special case of one-node states} \label{genqm}
\noindent 
The simplest example we can provide is the one that concerns the quantum mechanics of a one-particle system. This would correspond within the DNN/TQNN correspondence to an architecture where the boundary states consist of a single node. Hidden (DNN) and quantum (TQNN) layers differ in this picture by their cardinality, i.e. the number of copies interpolating among the boundary (one-node) states, and their size, i.e. the number of nodes at each intermediate layer. This difference is substantially prompted by the fact that classical DNNs have a fixed architecture, which corresponds to a single path in the Feynman formulation, while TQNNs superpose all the possible paths. Hidden layers in DNNs are indeed finite in number of copies and finite in size, while quantum layers can be also infinite in number of copies and in size. DNNs would then account for a finite sum over a finite set of intermediate steps, while quantum mechanics realizes a sum over infinite possible quantum states. Furthermore, DNNs assign probability weights among the edges connecting one-node states to nodes of the hidden layers, while TQNNs assign to the internal edges probability amplitudes, i.e. complex numbers the absolute value square of which are probabilities. The infinite number of quantum layers between input and output layers arises from the fact that any TQFT used to define a TQNN is invariant under triangulation changes, i.e. it is invariant under Pachner moves, and therefore arbitrary refinements of the simplicial decomposition between input and output are all taken into account by the TQFT. A related perspective concerns the Matveev-Piergallini moves, as described in \cite{Kauffman-Lins} for the quantum group $\mathcal U_q(\mathfrak{sl}_2)$.

Representing then boundary states with one-particle quantum states of quantum mechanics, we can associate 
\begin{eqnarray}
&|{\rm in} \rangle \rightarrow |x_{\rm in}, t_{\rm in} \rangle \,, \nonumber\\
&|{\rm out} \rangle \rightarrow |x_{\rm out}, t_{\rm out} \rangle \,,
\end{eqnarray}
where the Cartesian coordinate $x$ labels the particle position (in a one-dimensional linear space) and $t$ denotes the time parameter deployed to follow its dynamical evolution.

Suppose now to partition $t\in [t_{\rm in}, t_{\rm out}]$ in $N-1$ steps of interval $\Delta t= (t_{\rm out} - t_{\rm in})/(N-1)$. 
%Within DNNs, $N$ remains finite, 
While in TQNNs we take the limit $N\rightarrow \infty$, when considering the semi-classical limit for DNNs $N$ is fixed to be a finite number. Now, from the perspective of TQNNs, we estimate the classifier $\mathcal{Z}_{\ell} (x_{\rm in}, x_{\rm out} )$ as the partition function of the one-particle system, with $x_{\rm in}$ and $x_{\rm out}$ labelling the boundary states, and $\ell$ labelling the infinite set of all the possible paths connecting $x_{\rm in}$ to $x_{\rm out}$, as determined by the partition function of the underlying TQFT. This means that  
\begin{equation}
\mathcal{Z}_{\ell} (x_{\rm in}, x_{\rm out} )=\langle x_{\rm out}, t_{\rm out} |  x_{\rm in}, t_{\rm in} \rangle = \langle x_{\rm N}, t_{\rm N} |  x_{\rm 1}, t_{\rm 1} \rangle\,.
\end{equation}
  
Now we can decompose $\mathcal{Z}_{\ell} (x_{\rm in}, x_{\rm out} )$ into the product of the corresponding intermediate steps (using the functoriality of the theory) and get
\begin{eqnarray}
\langle x_{\rm N}, t_{\rm N} |  x_{\rm 1}, t_{\rm 1} \rangle \!\!\!\!\!\!&=&\!\!\!\!\!\! \int \!dx_{N-1}\! \int dx_{N-2} \dots  \int \!dx_{2}\, \times   \\
&\phantom{a}& \langle x_{\rm N}, t_{\rm N} |  x_{\rm N-1}, t_{\rm N-1} \rangle  \times 
\nonumber \\
&\phantom{a}&  \langle x_{\rm N-1}, t_{\rm N-1} |  x_{\rm N-2}, t_{\rm N-2} \rangle \dots \langle x_{\rm 2}, t_{\rm 2} |  x_{\rm 1}, t_{\rm 1} \rangle\,. \nonumber 
\end{eqnarray}
Within each interval of time $\Delta t$, we consider the Feynman amplitude  
\begin{equation} \label{Feynman}
\langle x_{k}, t_{k} |  x_{k-1}, t_{k-1} \rangle = \frac{1}{w(\Delta t)}\, e^{\imath S(k, k-1)}\,,
\end{equation}
where $\imath=\sqrt{-1}$ denotes the imaginary unit and 
\begin{equation} \label{action}
S(k, k-1)\equiv \int_{t_{k-1}}^{t_{k}} dt\,  L_{\rm classical}(x,\dot{x})
\end{equation}
is the classical action evaluated in the $k$-th time interval $\Delta t$, with $k=1, \dots N-1$.

Notice that the generic $k$-th amplitude has the meaning of a propagator from the point $\{x_{\rm k}, t_{\rm k}\}$ to the point $\{x_{\rm k-1}, t_{\rm k-1}\}$, since the generic intermediate states $|x_{\rm k}, t_{\rm k} \rangle$ and $|x_{\rm k-1}, t_{\rm k-1} \rangle$ evolve according to the Schr\"odinger picture, i.e. 
\begin{eqnarray}
\langle x_{\rm k}, t_{\rm k} |  x_{\rm k-1}, t_{\rm k-1} \rangle &=& \langle x_{\rm k}| e^{-\frac{\imath}{\hbar} H \Delta t } |x_{\rm k-1} \rangle \nonumber \\
&=& K(x_{\rm k}, t_{\rm k}; x_{\rm k-1}, t_{\rm k-1})\,.
\end{eqnarray}
By construction, the propagator $K(x_{\rm k}, t_{\rm k}; x_{\rm k-1}, t_{\rm k-1})$ satisfies the Schr\"odinger time-dependent wave equation, and the property   
\begin{equation}
\lim_{t_{\rm k} \rightarrow t_{\rm k-1}} K(x_{\rm k}, t_{\rm k}; x_{\rm k-1}, t_{\rm k-1})=\delta(x_{\rm k} - x_{\rm k-1})\,.
\end{equation}
The propagator hence determined is nothing but the Green's function of the time-dependent Schr\"odinger wave equation, i.e.
\begin{eqnarray} \label{SCH}
&\left[ H(x_{\rm k})- \imath \hbar \frac{\partial}{\partial t_{\rm k}}\right] K(x_{\rm k}, t_{\rm k}; x_{\rm k-1}, t_{\rm k-1})= \nonumber\\
&\delta(x_{\rm k} - x_{\rm k-1}) \delta(t_{\rm k} - t_{\rm k-1})\,,
\end{eqnarray}
where $H(x_{\rm k})$ is the differential representation of the Hamiltonian operator in $x_{\rm k}$, and with boundary condition 
\begin{eqnarray}
K(x_{\rm k}, t_{\rm k}; x_{\rm k-1}, t_{\rm k-1})= 0\,, \qquad \forall t \notin [t_{\rm k-1}, t_{\rm k}]\,.
\end{eqnarray}

In \eqref{Feynman}, the factor in front of the exponential can only depend on the time interval $\Delta t$. Being independent from the potential to which the particle is subjected, it can be estimated by calculating from \eqref{SCH} the propagator of a non-relativistic free particle, with Hamiltonian $H=\frac{1}{2}m \dot{x}^2= -\frac{\hbar^2}{2m} \nabla^2$, hence finding the expression 
\begin{eqnarray}
\frac{1}{w(\Delta t)}=\sqrt{\frac{m}{2 \pi \imath \hbar \Delta t}}
\,.
\end{eqnarray}
We can now consider the limit for which the partition acquires an infinite amount of ``filters'', and hence the time interval $\Delta t$ shrinks to zero. Within this limit, each $k$-th amplitude will contribute according to 
\begin{eqnarray}
\langle x_{\rm k}, t_{\rm k} |  x_{\rm k-1}, t_{\rm k-1} \rangle = \sqrt{\frac{m}{2 \pi \imath \hbar \Delta t}} \, e^{\frac{\imath}{\hbar} S(k, k-1)}\,,
\end{eqnarray}
providing as a final expression for the transition from $| {\rm in}\rangle$ to $| {\rm out}\rangle$ the relation 
\begin{eqnarray}
\langle x_{\rm N}, t_{\rm N} |  x_{\rm 1}, t_{\rm 1} \rangle &=& \lim_{N \rightarrow{\infty} } \left(\frac{m}{2 \pi \imath \hbar \Delta t}\right)^{\frac{N-1}{2}} \,\int dx_{N-1} \int dx_{N-2}  \nonumber\\ 
&\phantom{a}& \dots \int dx_{2} \prod_{k=2}^N \, e^{\frac{\imath}{\hbar} S(k, k-1)}\,.
\end{eqnarray}
By the definition of $S(k, k-1)$ in \eqref{action}, and denoting the sum over the paths as $\mathcal{D}[x(t)]$, namely 
\begin{eqnarray}
\int_{x_{1}}^{x_{\rm N}} \mathcal{D}[x(t)]=\lim_{N \rightarrow{\infty} } \left(\frac{m}{2 \pi \imath \hbar \Delta t}\right)^{\frac{N-1}{2}} \!\!\!\!\!\!\!\!\!\!\!\!&\phantom{a}& \,\int dx_{N-1} \times \\
&\phantom{a}& \int dx_{N-2}  
 \dots \int dx_{2}  \,,  \nonumber
\end{eqnarray}
we find the celebrated expression for the Feynman path-integral 
\begin{eqnarray}
\langle x_{\rm N}, t_{\rm N} |  x_{\rm 1}, t_{\rm 1} \rangle = \int_{x_{1}}^{x_{\rm N}} \mathcal{D}[x(t)] \, e^{\frac{\imath}{\hbar} \int_{t_1}^{t_N} L_{\rm classical}(x,\dot{x}) }\,.
\end{eqnarray}

When a free particle is considered, one can observe that the semi-classical limit corresponds to the minimisation of the action, hence to imposing the stationarity of the norm of $X=p/\sqrt{2m}$. Taking into account relativistic invariance, in a $d+1$-dimensional space-time manifold, the vector $\vec{X}=\vec{p}/\sqrt{2m}$ will turn out to be $d$-dimensional. Thus the dimensionality of the vector $\vec{X}$ equals the space dimension of the ambient space-time manifold. 

The interpretation we provide here is straightforwardly preserved every time we consider a Lagrangian that is quadratic in the configuration variables and their momenta. More in general, we can resort to a symplectic geometry analysis to identify in full generality the norm of the vector $\vec{X}$ in terms of an Hermitian inner product. This latter is generated by the symplectic structure associated to the manifold --- see e.g. Refs.~\cite{CW,C1,C2} --- and corresponds to the norm of the action of the space-time translation generator applied to the configuration fields of the system. In fact, when the Lagrangian presents higher order terms, one can proceed by perturbing the path integral as shown for example in \cite{sawon2006perturbative} for Chern-Simons theory. This allows to compute higher order terms in the perturbation as described above. Moreover, this paves the way to further parametrisations that can be learned during training in the form of topological charges. This situation, albeit very interesting, will not be considered explicitly in the present article.

Notice furthermore that once manifolds with Lorentzian signature are taken into account, the extremisation of the action would not correspond to a minimisation of $\vec{X}$, because the system is hyperbolic. In this latter case, the extremisation of the action on a hyperbolic manifold provides the classical trajectories/geodesics of the systems, namely the classical paths that instantiate generalisation within the DNNs framework.

Therefore the extremisation of the weights, whose role in generalisation has been commented in Ref.~\cite{30}, is replaced in this picture by the minimisation of the action. This latter in turn corresponds to maximise the probability amplitudes in the path integral formulation of a classifier.

\begin{figure}[h!]
\centering
\begin{center}
\begin{tikzpicture}
\draw[line width=0.35mm] (5,3) ..controls(4.5,4) and (1.5,3).. (0,0);
\draw[line width=0.35mm,dashed] (5,3) ..controls(4.7,4.25) and (1.25,3.5).. (0,0);
\draw[line width=0.35mm,dashed] (5,3) ..controls(4.7-.25,4.25-.25) and (1.25-.25,3.5-.25).. (0,0);
\draw[line width=0.35mm,dashed] (5,3) ..controls(4.5-.25,4-.25) and (1.5-.25,3-.25).. (0,0);
\draw[line width=0.35mm,dashed] (5,3) ..controls(4.5-.25,4-.5) and (1.5-.25,3-.5).. (0,0);
\node (a) at (5.5,3) {$(x_1,t_1)$};
\node (a) at (-.5,0) {$(x_0,t_0)$};
\end{tikzpicture}
\end{center}
\vspace{-0.6cm}        
\caption{\footnotesize Predominant paths in the $h \rightarrow 0$ limit.}
\label{fig:hlimit}
\end{figure}

We are now in the position to express our thesis on the generalisation process.
The partition function that determines the TQFT used by the TQNN utilises a superposition of all colorings of the input/output pairs. The intermediate states amplitudes, as shown above, provide the single probabilities (upon taking moduli squared) of the single transitions of color configurations. Furthermore, they provide the summands of the partition function, as well as the single probabilities (upon taking moduli squared) of the single transitions of color configurations. Optimisation here shows the dominant terms in the superposition function, according to the given underlying ground truth. Thus generalisation emerges as an artifact of the semiclassical limit: DNNs structures are only able to represents fixed 2-skeletons within the 2-complex evolution of TQNNs. These are nevertheless the most dominant contributions to the path integral for the TQNNs, which is in general realised by the topological quantum neural 2-complexes (TQN2Cs), summing over all the quantum histories.
This simple case with boundary states with single nodes is generalised directly to the case of spin-networks with higher number of nodes, where the reasoning still holds true. In fact, this formulation was used in \cite{TQNN}. 

\subsection{Generalisation for TQNNs} \label{genTQNN}
\noindent 
We have so far addressed the process of generalisation for DNNs as the semiclassical limit of TQNNs, with particular emphasis on the case of single-node boundary states. However, a more general question naturally arises, regarding the notion of generalisation for TQNNs. In other words, we ask (and provide an answer to) the question of what a TQNN learns, and therefore what the generalisation procedure looks like in the quantum case.

Although the notion of TQNN seems to share similarities with graph neural networks (GNNs), we observe that the two methods are substantially different. In fact, GNNs are determined by a fixed geometry and, in addition, such geometry determines the way information is processed through the learning procedure (message passing). On the contrary, TQNNs have no fixed geometric structure, but they are applied through their defining functorial rules on graphs that correspond to spin-networks, and therefore elements of the boundary Hilbert space, via some pre-determined assignment (see \cite{TQNN} Section~4 for such a concrete correspondence). Information from the boundary states is then processed according to the defining rules of the given choice of TQFT --- quantum $\mathfrak {sl}_2$ in the case discussed in detail in \cite{TQNN}. A TQFT is supported on the bulk manifold and it is independent on the triangulation used. It follows that the information contained in the boundary is processed according to methods that are not determined by the geometric structure of the supporting boundary graph, as it happens in the case of message passing for GNNs. This is a fundamental perspective that TQNNs leverage to generalise.

TQNNs evolve according to the dynamics dictated by TQFTs. Transition amplitudes calculated according to TQFTs instantiate classifier rules that interconnect TQNN states. As in \cite{TQNN}, we focus on $BF$-extended theories, which is a theoretical framework general enough to encode Yang-Mills gauge theories as well as effective field theories. Specifically, TQNNs are states of the kinematical Hilbert space of TQFTs. Their topological features are easily captured once their expansion on the spin-networks basis is taken into account. Nonetheless, an equivalent expansion in the multi-loop basis is also possible \cite{rovelli1995spin}, which renders less evident the connectivity of graphs and thus (some of) their topological features. Such expansion is obtained by unraveling the Jones-Wenzl symmetrizer (projector) placed on each edge of the input boundary, following the definition of spin-network state. For our purposes, it suffices to consider the expansion of TQNNs on the spin-networks basis, found e.g. in \cite{baez1996spin}.

We consider a manifold $\mathcal{M}$, 
and a submanifold $\mathcal S$ such that $\mathcal M = \mathcal M_0 \cup_{\mathcal S} \mathcal M_1$, for submanifolds $\mathcal M_i$, $i=0,1,$ with $\partial \mathcal M_i = \mathcal S$. The one-complexes (graphs) $\gamma$ are then considered as embedded in $\mathcal S$. A generic TQNN state can be expanded on the elements of the spin-network basis, each one being supported on a generic graph $\gamma \in S$. In turn, each spin-network state is defined as a triple $\Psi=(\gamma_\Psi, \rho, \iota)$ consisting of a graph $\gamma_\Psi\in S$, an irreducible representation $\rho_l$ of $G$ for each link $\gamma_i\in \gamma_\Psi$, an intertwiner $\iota_n$ for each node $n$ such that
\begin{equation}
\iota_n: \rho_{l_1} \otimes \dots \otimes  \rho_{l_n} \rightarrow \rho_{l_1'} \otimes \dots \otimes  \rho_{l_n'}\,,    
\end{equation}
where the links incoming into the node $n$ have been denoted as $l_1, \dots l_n$, and the links outgoing from the node $n$ have been denoted with $l_1', \dots l_n'$.

Without loss of generality, we may directly focus on a generic spin-network state $\Psi$. 
Suppose now that $\Psi$ is the spin-network state associated to some boundary initial state derived from a data-point, namely 
\begin{equation}
|{\rm in} \rangle \rightarrow |\Psi \rangle\,,
\end{equation}
where the correspondence is determined according to some given rule as in Section~4 of \cite{TQNN}, for example.
Similarly, introduce the boundary final state $\Phi$, i.e.
\begin{equation}
|{\rm out} \rangle \rightarrow |\Phi \rangle\,. 
\end{equation}
Generic TQNN states $\Psi$ and $\Phi$ encode two types of data: topological data, namely the connectivity of the graphs $\gamma_\Psi$ and $\gamma_\Phi$; metric data, corresponding to the assignments of quantum numbers to the links and nodes of the graph $\gamma_\Psi$ and $\gamma_\Phi$, namely the irreducible representations $\rho_l$ for each link $l\in \gamma_\Psi$ or $l\in \gamma_\Phi$, and the intertwiner quantum number $\iota_n$ for each node $n\in \gamma_\Psi$ or $n\in \gamma_\Phi$. Both the topological and metric data that are encoded in $\Psi$ and $\Phi$ are processed by the classifier $\mathcal{Z}$. This latter is nothing but the matrix element of the quantum evolution operator --- in the path integral representation --- that associates the initial state $|{\rm in} \rangle = |\Psi \rangle$ to the final state $|{\rm out } \rangle = |\Phi \rangle$. The action of the evolution operator/classifier on the boundary states amounts to the assignment of the transition amplitude $\langle \Phi | \Psi\rangle_{\rm phys}$. Here, the subscript ``phys'' is reminding us that the amplitude is calculated in terms of the dynamics of the specific TQFT taken into account, thus it differs from the scalar product $\langle \Phi | \Psi\rangle$ calculated in the kinematical Hilbert spaces to which the TQNN states $\Psi$ and $\Phi$ belong. In practice, the transition amplitude is one of the summands appearing in the partition function that defines the chosen TQFT. 

Once a specific TQFT is selected, and a gauge group $G$ is fixed, the states of the TQNN can be represented as cylindrical functionals 
\begin{equation}
\Psi(H_l):=\langle H_l| \Psi\rangle
\end{equation}
that depend on the holonomies $H_l\in G$ along the links $l\in \gamma_\Psi$. Holonomies are group elements of $G$ that realize the parallel transport along a path $\gamma  \subset S \subset \mathcal{M}$, with respect to the connection $A$ over the principle $G$-bundle. The graph $\gamma_\Psi$ here represents a discretisation of the underlying manifold, and its links represent small paths in the manifold --- cf. the notion of edge in lattice gauge theory. Denoting path-ordering with $P$, the parallel transport of a vector in a representation $\rho$ of $G$ reads 
\begin{equation}
H_l= P e^{ \int_l A_a \tau^a}\,,    
\end{equation}
with $a$ an index in the adjoint representation of $G$ and $\tau^a$ a representation of the generators of $G$.

The classifier quantum amplitudes can be derived for an extended $BF$-theory over a $G$-bundle. We consider a local trivialisation on $\mathcal{M}_d$, and denote the curvature of the $G$-connection $A$, which is a $\mathfrak{g}$-valued 1-form, with a $\mathfrak{g}$-valued 2-form $F$, $\mathfrak{g}$ standing for the Lie algebra of $G$. A frame field $B$ can be introduced, as the field conjugated to $A$, and such that the symplectic structure is fulfilled
\begin{equation}
\omega((\delta A, \delta B), (\delta A', \delta B') )= \int _S \langle \delta A \wedge \delta B'- \delta A' \wedge \delta B \rangle\,,
\end{equation}
where the conjugated fields have been restricted on an initial (in time, for the Lorentzian case) slice $\{0\} \times S$ of $\mathcal{M}_4$, and we have denoted the trace over the internal indices with $\langle \dots \rangle$. Notice that the frame field $B$ is a $\mathfrak{g}$-valued $2$-form. This allows to write consistently the action of the $BF$-extended theory over either a Riemannian or a Lorentzian $4$-dimensional manifold $\mathcal{M}_{4}$ as 
\begin{eqnarray}
\label{BF-extended}
\mathcal{S}_{\rm BF}^{\rm ext.}[A,B]\! =\!\! \int_{\mathcal{M}_4} \!\! \! \langle B\wedge F + \lambda_1 B\wedge B + \lambda_2 B\wedge \star B\rangle\,, \ \ \ \
\end{eqnarray}
with $\lambda_1, \lambda_2\in \mathbb{R}$ bare coupling-constants. For $\lambda_1\neq 0$ and $\lambda_2=0$, the theory is still topological, and corresponds to the Crane-Yetter model, whose quantisation involves recoupling theory of quantum groups --- see e.g. the notable example that corresponds to $G=SU(2)$.
For $\lambda_1= 0$ and $\lambda_2\neq0$, the theory is non-topological and corresponds on-shell to a Yang-Mills action with internal gauge group $G$. Furthermore, when the Lorentzian case with $G=SL(2, \mathbb{C})$ is taken into account, and $\lambda_1$ is promoted to a multiplet of scalar fields, with a pair of symmetric indices that are in the adjoint representation of $SL(2, \mathbb{C})$, one can recover the Einstein-Hilbert-Holst action of gravity for $4$-dimensional space-time \cite{fgm:22}.

For the sake of simplicity, we focus on the topological realization of the $BF$-theory that corresponds to selecting $\lambda_1=\lambda_2=0$. The equations of motions then read
\begin{equation}\label{EOM2}
 F=0\,,  \qquad \qquad   d_A B=0\,,
 \,
\end{equation}
where $d_A$ denotes the covariant derivative with respect to the $G$-connection $A$. 

The classifier evolution is easily recovered in terms of the instantiation of the curvature constraint $F=0$ in a physical projector $\mathcal{P}$. In turn, the implementation of the physical projector $\mathcal{P}$  was discussed in \cite{Noui-Perez}, where it was shown how to make sense of the formal expression
\begin{equation} \label{phys-pro}
\mathcal{P}=\int \mathcal{D} N \exp (\imath \int_S \langle N \, \widehat{F}\rangle  )
\,,
\end{equation}
with $N$ a $\mathfrak{g}$-valued Lagrangian multiplier. 

In particular, the curvature constraint $F=0$ can be implemented in the physical amplitudes among spin-network states according to
\begin{equation} \label{inn-pro-phys}
\langle \Psi , \Phi \rangle_{\rm phys} = \langle \mathcal{P} \Psi , \Phi \rangle\,.
\end{equation}

A local patch $\Sigma \in S$ can be considered that is provided with cellular decomposition composed of squares of infinitesimal coordinates length $\delta$. The regularised curvature constraint then reads
\begin{equation}
F[N]= \int_\Sigma \langle N F(A) \rangle = \lim_{\delta \rightarrow 0} \sum_{p^j} \delta^2 \langle N_{p^j} F_{p^j} \rangle\,,
\end{equation}
in which ${p^j}$ labels the $j$-th plaquette and $N_{p^j}$ is the discretisation of the $\mathfrak{g}$-valued Lagrangian multiplier $N$, evaluated at an interior point of the plaquette ${p^j}$.

The holonomy $H_{p^j}[A]$ around the plaquette ${p^j}$ is a $G$ group-element that casts
\begin{equation}
H_{p^j}[A]= 1\!\!1 + \delta^2 F_{p^j}(A) + \mathcal{O}(\delta^2)
\,,
\end{equation}
which implies
\begin{equation} \label{smeared-curve}
F[N]= \int_\Sigma \langle N F(A) \rangle = \lim_{\delta \rightarrow 0} \sum_{p^j} \delta^2 \langle N_{p^j} H_{p^j}[A] \rangle\,.
\end{equation}
The regularised expression for the action of the physical projection operator immediately follows. This enters the physical scalar product of spin-network states:
\begin{eqnarray} \label{reg-inn-pro}
\langle \Psi , \Phi \rangle_{\rm phys} &=& \lim_{\delta \rightarrow 0} \langle \prod_{p^{j}} \int \mathcal{D} N_{p^{j}} \exp(\iota \langle
N_{p^{j}} \widehat{H}_{p^{j}} \rangle)  \Psi, \Phi  \rangle \nonumber\\
&=& \lim_{\delta \rightarrow 0} \langle \prod_{p^{j}} \delta(H_{p^j}) \Psi, \Phi \rangle\,.
\end{eqnarray}

%%%%%%%%%%%%%
Having reminded the structures of TQNNs and their TQFT evolution in terms of TQN2C classifier, we can now address the problem of generalisation in this extended theoretical framework. We may assume that the size of the training data is sufficient to select or, better, to learn specific paths in the boundary graphs and bulk 2-complexes within the most general available TQNN architecture. These paths are characterised by different types of associated non-perturbative topological charges, which in turn provide the sub-structures involved in the generalisation process, as a subset supported on general 2-complexes. More concretely, these correspond to summands in the partition function of the TQFT that is used to define the TQNN, and determine the dominating contributions to the topological invariants. In other words, certain intrinsic algebro-geometric features emerge that characterise the learning of the TQNN.  

The BF-extended formulation provided by \eqref{BF-extended} encode Yang-Mills theories over a $G$-bundle for the choice $\lambda_1=0$ and $\lambda_2=2 g^2_{\rm YM}$. Thus the optimisation of the classifier will correspond to the minimisation of the classical Yang-Mills action, namely  
\begin{eqnarray}
\mathcal{S}_{\rm YM}=  \frac{1}{2 g^2_{\rm YM}} \int_{\mathcal{M}_4}\langle F\wedge \star F \rangle\,.
\end{eqnarray}
Cast in terms of the spacetime components, with $a$ denoting the indices of the adjoint-representation of the algebra $\mathfrak{g}$, the Lagrangian density $F_{\ \mu \nu}^a F^{a \ \mu \nu}$ turns out to be proportional to the norm of a tensor on a hyperbolic manifold with Lorentzian signature. 

Path selection is attained by minimisation of the classical action and extremisation of the path-integral formulation of the classifier, which is naturally realised in the semi-classical limit, in a way that is reminiscent of the free energy principle formalism \cite{friston:10, friston:13, friston:19, ffgl:22}.

The topological charges that are switched on over the learning process and correspond to the configurations that extremise the action, together with the corresponding metric properties, implement effectively the generalisation process. In this sense, our approach is expected to provide a solution to the problem as raised by Zhang et al \cite{zhang2017}. In particular: 

\begin{itemize}

    \item 
The randomisation of the labels of a TQNN state will not induce overfitting as a consequence of the encoding of information achieved by the TQNN through the topological features and topological invariants. The quantum nature of the TQNN will induce fluctuations around values of the parameters to be estimated. Nonetheless, these fluctuations are small in the semi-classical limit.

\item 
However, a DNN architecture will be trapped into an overfitting regime until memorising the training examples by brute force, since by definition of DNNs the training error vanishes --- the variance for the $j$ scale as $1/\sqrt{\bar{j}}$. In other words, associating a DNN to a set of spin-networks evaluated into coherent group elements, the corresponding training error is zero.
    
\end{itemize}

Thus, brute-force learning is not possible for a TQNN, since the topological information of the input/output states would automatically make the transition amplitude vanish, most of the time, with random labelling.

\section{Perceptron as a semi-classical limit of TQNN}\label{perce}
\noindent 
We address in this section the case of the perceptron, considering it as the semi-classical limit of TQNN.

Within the large $j_{ab}$-spin limit, i.e. in the semi-classical limit, the transition amplitudes among spin-network states become
\begin{eqnarray}
\mathcal{A}_{\prod_{ab} H_{ab}, \vec{w}} &=& \langle \psi_{\Gamma_{\vec{\chi},H_{ab} }} | \psi_{\Gamma_{\vec{w}, j_{\vec{w}}, \iota_n} }\rangle   \nonumber \\
&=& 
\prod_{ab} \Delta_{j_{ab}} e^{-\frac{(j_{ab}-\bar{j}_{ab})^2}{2 \sigma^2_{ab}}}e^{-\iota \xi_{ab}j_{ab}},
\label{eq:amplitude}
\end{eqnarray}
where $H_{ab}$ denote SL(2, $\mathbb{C}$) elements data in the asymptotic limit, $\Delta_{j_{ab}}$ denotes the dimension of the irreducible representation, labeled by the spin $j_{ab}$, $\sigma_{ab} \equiv 1/(2t_{ab})$ is related to the diffusion time $t_{ab}$ and $\xi_{ab}$ is the dihedral angle between input and output. \\
The probability associated with the amplitude in Eq.~\eqref{eq:amplitude} is not normalised. In order to obtain a normalised probability, we proceed to define
\begin{equation}
|\bar{\mathcal{A}}_{\prod_{ab} H_{ab}, \vec{w}}|^2 = \frac{|\mathcal{A}_{\prod_i H_{ab}, \vec{w}}|^2}{\mathrm{max}(|\prod_{ab}\Delta_{j_{ab}}|^2,|\prod_{ab}\Delta_{\bar{j}_{ab}}|^2)}\,.
\label{eq:prob}
\end{equation}
The probability amplitude within Eq.~\eqref{eq:amplitude} is at the base of a novel definition of perceptron that solves the task of image classification by categorising their topological features. We test this idea on a toy model, the handwritten digits of the MNIST dataset. 

\subsection{Preprocessing and Architecture}
\label{sec:perceptron}
\begin{figure}[h!]
  \centering
    \includegraphics[width=\linewidth, scale = 1]{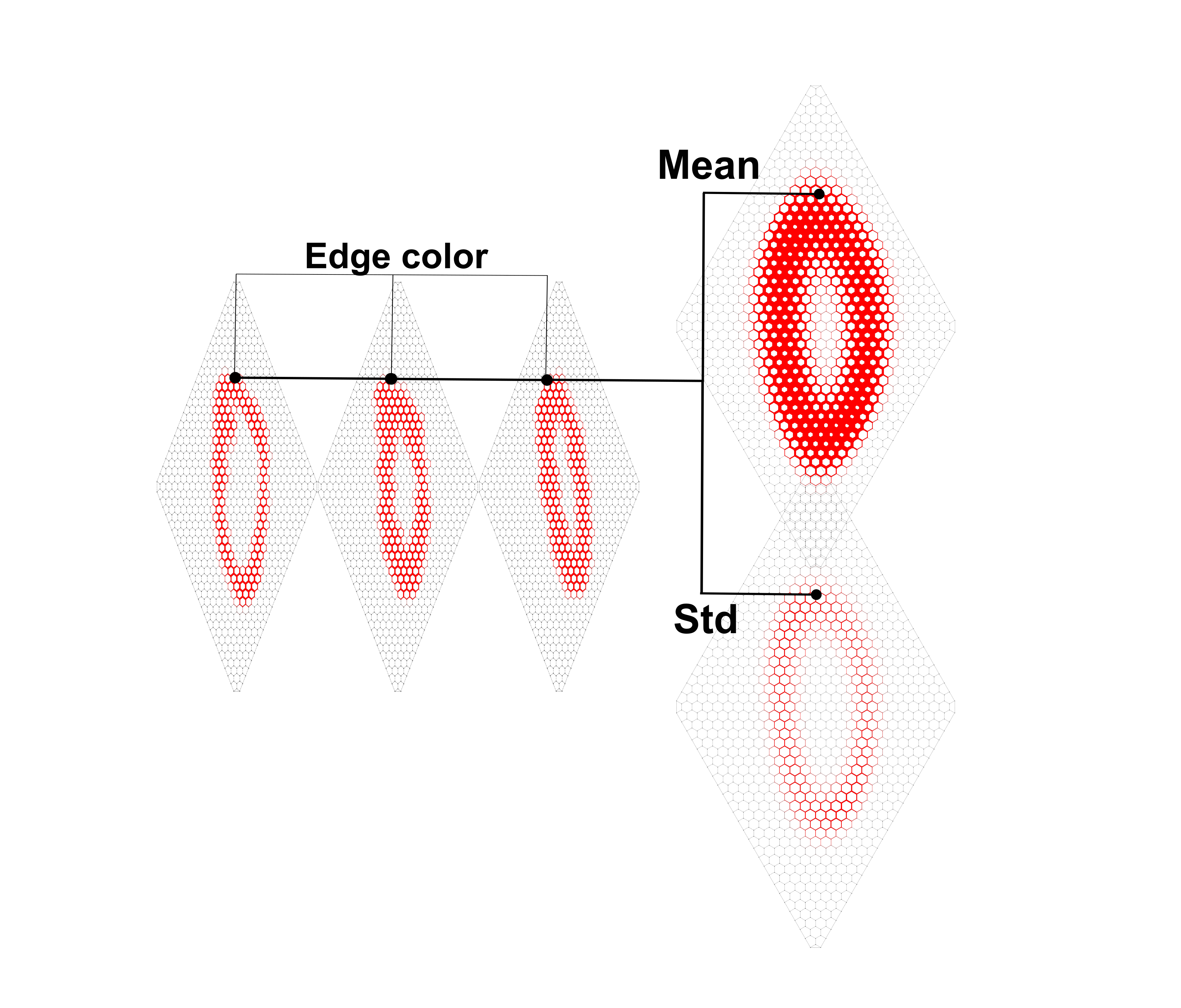}
    \caption{Outline of the proposed algorithm. Select all digits with label 0 in the training set and evaluate the mean and standard deviation. This defines the output state of the semiclassical theory.}
    \label{fig:perceptron}
\end{figure}

The MNIST dataset consists of 60000 images of handwritten digits of shape $28 \times 28$. We first divide the dataset into training and testing sets with a ratio 80-20. Then we use the algorithm introduced in \cite{lulli2023exact} to define the spin-network states associated with the images of the MNIST dataset. We obtain in this way 2349 edge colors for each image in the dataset. These are the parameters and the inputs of the perceptron. For numerical reasons, we apply a global normalisation to the colors of the spin-networks, so that we have colors within the range [0,1]. Eq.~\eqref{eq:amplitude} has three parameters that we need to determine, $\bar{j}_{ab}$, $\sigma_{ab}$ and $\xi_{ab}$, through the use of the training dataset. Differently than for the standard perceptron, we do not implement any optimisation process based on gradient descent, nor any other of its variants. 
Parameters have been chosen in analogy with geometric theories in physics, and according to computational learning theory.
%%
%%
%%\textcolor{red}{Parameters have been chosen from physical theory.[AM: I did not get this point. What does it mean?Maybe: Parameters have been chosen in analogy with geometric theories in physics. T: I agree with you]} \textcolor{purple}{[KJ: I think it should mean "computational learning theory" as we mentioned it as machine learning without machine learning discussion last time.]} 
%%
%%
For this purpose, we divide the spin-networks associated with the digits into groups corresponding to their labels. For each of the 10 groups, we define 
\begin{align}
\langle\bar{j}_{ab}\rangle_{k} &= \frac{1}{N_k}  \sum_{l=1}^{N_k} \bar{j}_{ab, l}, \\
\sigma_{ab, k}^2 &= \langle\bar{j}^2_{ab}\rangle_{k} - \langle\bar{j}_{ab}\rangle_{k}^2, \\
\xi_{ab, k} &= 0
\end{align}
where $k$ indicates the label of the digits and $N_k$ is the number of digits with label $k$. In this case, the dihedral angle between input and output is set to zero, since the images are defined on the 2D plane. The diagram is shown in Fig.~\ref{fig:perceptron}. \\

For each set of parameters $\{\bar{j}_{ab}, \sigma_{ab}, \xi_{ab} \}_k$ and test spin-networks' asymptotic parameters $H_{\mathrm{test},\,ab}$, we evaluate the probability in Eq. \eqref{eq:prob}. The $argmax_k \left( \log(|\bar{\mathcal{A}}_{\prod_{ab} H_{\mathrm{test},ab}; \{\bar{j}_{ab}, \sigma_{ab}, \xi_{ab} \}_k}|^2) \right)$ then provides the classification of the example. The use of the logarithm has been considered for computational reasons, in order to avoid underflow. We expect that the needed normalisations and use of logarithms as described above affect the result negatively in a significant way. However, we mention that an approach to such implementation that would be devoid of additional computational errors would most likely be a quantum computer implementation of our approach that uses material-science-based implementations of spin-network transitions. Such technology is not yet available to us.

%\subsection{Robustness test}
%We evaluate the robustness of the perceptron defined in Sec. \ref{sec:perceptron}. We first study the impact that the choice of hyperparameters has on the performance of the algorithm. For this purpose, we use the k-fold cross-validation method to understand how the choice of parameter set $\{\bar{j}_{ab}, \sigma_{ab}, \xi_{ab} \}_k$ affects the classification. The number of k-folds chosen is 100, and each parameter set is evaluated using 43 train examples for each label. Leaving the test set unchanged, the transition probability in Eq. \ref{eq:prob} is then evaluated, for each batch and label. \\
%We then study what happens by changing the value of the standard deviation $\sigma_{ab,k}$, and the value of the dihedral angle $\xi_{ab,k}$. The dihedral angle will be chosen from a Gaussian distribution for each label $k$. We study the change in scores as the mean and standard deviation of the distribution change. \\
%Another important test is the robustness to noise of the TQFT-based algorithm compared to traditional perceptrons. The noise is sampled from a zero mean Gaussian distribution where the standard deviation is varied. The noise thus sampled is added to the test images, and the change in perceptrons performance is then monitored for each choice of standard deviation. 

\subsection{Results}
The performance of the TQFT perceptron in the semi-classical limit (with no training) is tested to show that it recovers two other architectures, the original perceptron with step function as the activation function and the perceptron with Softmax as activation function. Traditional perceptrons are implemented using the Python library \textit{Tensorflow}. To simulate the behavior of the step function we use a hard sigmoid, this is because the step function is not differentiable. The two perceptrons are trained for 10 epochs, using categorical cross-entropy as the loss function. While the aforementioned numerical approximations reduce the overall performance of the TQNN perceptron, we see that the results obtained are similar to the standard perceptron, therefore demonstrating that TQNNs in the semi-classical limit recover, without the need for training, standard trained perceptrons. 
The results are reported in Tab. \ref{tab:class}.

\begin{table}[!h]
    \centering
\begin{tabular}{cccccc} \toprule
    Label & Precision & Recall & F1-score & Acc & R2 - score \\ \midrule
     &&&&&\\
     \multicolumn{6}{c}{\textbf{TQFT}} \\ 
      &&&&&\\
    0 & 0.88 & 0.93 & 0.91 & & \\
    1 & 0.96 & 0.89 & 0.92 & &   \\
    2 & 0.86 & 0.89 & 0.87 & &  \\
    3 & 0.80 & 0.84 & 0.82 & &     \\ 
    4 & 0.85 & 0.86 & 0.85 & &  \\
    5 &  0.79 & 0.74 & 0.77 & &   \\
    6 & 0.93 & 0.93  & 0.93 & &  \\
    7 & 0.93 & 0.88  & 0.90 & &  \\ 
    8 & 0.79 & 0.80  & 0.80 & &  \\
    9 & 0.79 & 0.82  & 0.80 & &  \\
    &  &  & & 0.86 & 0.70 \\ \hline % \bottomrule
        &&&&&\\
    %  \multicolumn{6}{c}{\textbf{Gaussian}}\\ 
    %   &&&&&&\\
    %0 & 0.95 & 0.96 & 0.95 & & \\
    %1 & 0.91 & 0.99 & 0.95 & &   \\
    %2 & 0.88 & 0.91 & 0.90 & &  \\
    %3 & 0.92 & 0.86 & 0.89 & &     \\ 
    %4 & 0.95 & 0.87  & 0.91 & &  \\
    %5 & 0.83&  0.90 &  0.87 & &   \\
    %6 & 0.94  & 0.96   &0.95 & &  \\
    %7 & 0.97 & 0.88  & 0.92 & &  \\ 
    %8 & 0.94 & 0.82  & 0.88 & &  \\
    %9 & 0.81 & 0.94  & 0.87 & &  \\
    %&  &  & & 0.91 & 0.81 \\ \thickhline % \bottomrule
     &&&&&\\
          \multicolumn{6}{c}{\textbf{Softmax}}\\ 
           &&&&&\\
    0 & 0.93 & 0.98 & 0.95 & & \\
    1 & 0.98 & 0.96 & 0.97 & &   \\
    2 & 0.96 & 0.86 & 0.91 & &  \\
    3 & 0.87 & 0.91 & 0.89 & &     \\ 
    4 & 0.95 & 0.92  & 0.94 & &  \\
    5 & 0.96&  0.63 &  0.76 & &   \\
    6 & 0.93  & 0.97   &0.95 & &  \\
    7 & 0.97 & 0.88  & 0.92 & &  \\ 
    8 & 0.72 & 0.95  & 0.82 & &  \\
    9 & 0.84 & 0.91  & 0.88 & &  \\
    &  &  & & 0.90 & 0.80 \\ \hline % \bottomrule
     &&&&&\\
      \multicolumn{6}{c}{\textbf{Hard sigmoid}}\\ 
       &&&&&\\
    0 & 0.61 & 0.98 & 0.75 & & \\
    1 & 0.96 & 0.97 & 0.97 & &   \\
    2 &0.92 & 0.86 & 0.89 & &  \\
    3 & 0.83 & 0.90 & 0.87 & &     \\ 
    4 & 0.95 &0.89  &0.92 & &  \\
    5 & 0.93&  0.75 &  0.83 & &   \\
    6 & 0.96  &0.88   &0.92 & &  \\
    7 & 0.93 & 0.92  & 0.93 & &  \\ 
    8 & 0.90 & 0.69  & 0.78 & &  \\
    9 & 0.88 & 0.82  & 0.85 & &  \\
    &  &  & &  0.87 & 0.55 \\ \bottomrule

\end{tabular}
    \caption{Classification results for the test dataset. Comparison between the proposed architecture based on TQFT, original perceptron with step function, perceptron with Gaussian activation function and Softmax.}
    \label{tab:class}
\end{table}
The performances we obtain are in line with those recovered by using the original perceptron with activation given by the step function. With Softmax activation, the performance of the perceptron increases. 

\begin{figure}[h!]
\centering
\parbox{5cm}{
\includegraphics[width=5cm]{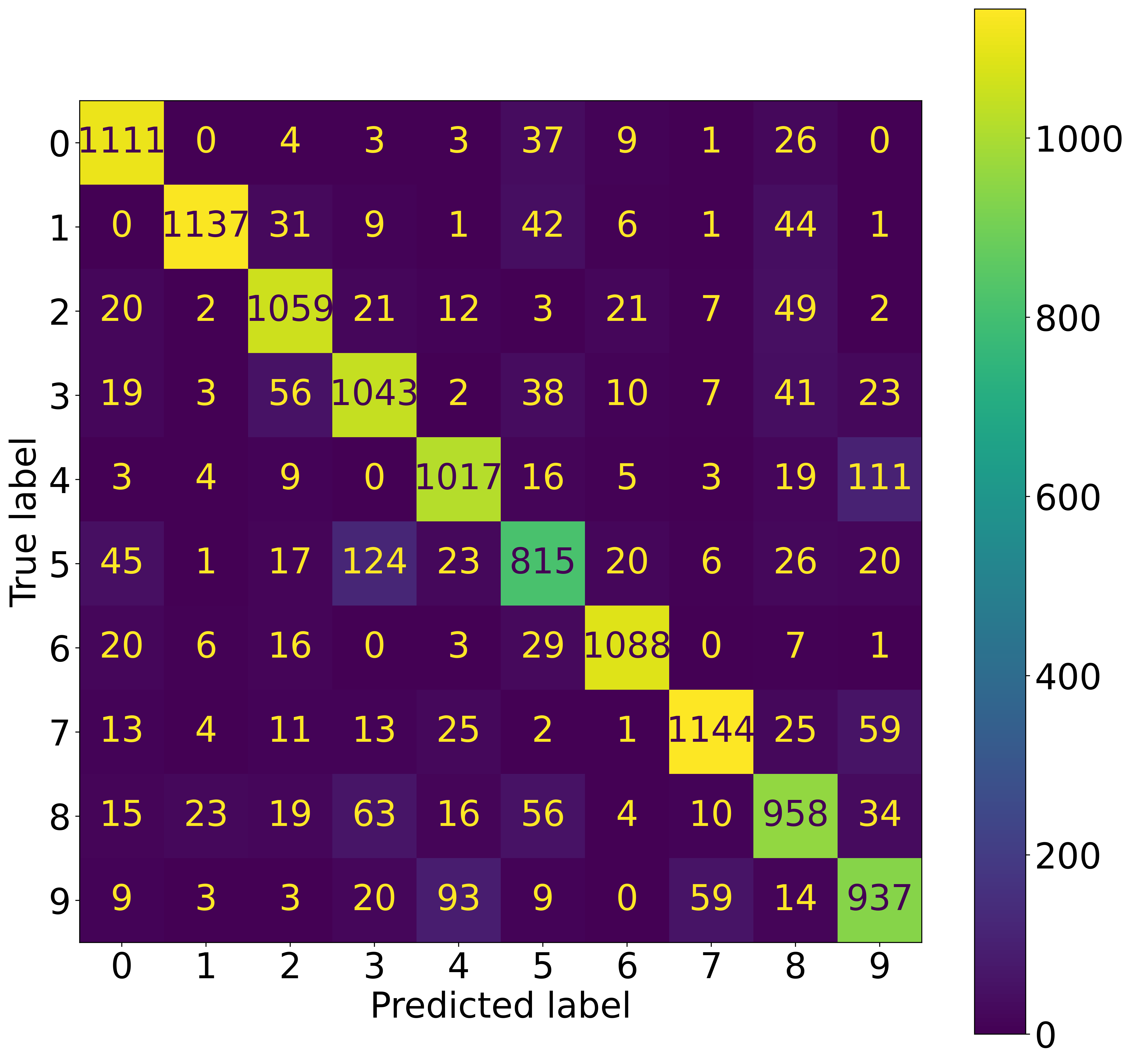}
\caption{Confusion matrix for the perceptron from TQFT}
\label{fig:cm1}}
\qquad
\begin{minipage}{5cm}
\includegraphics[width=5cm]{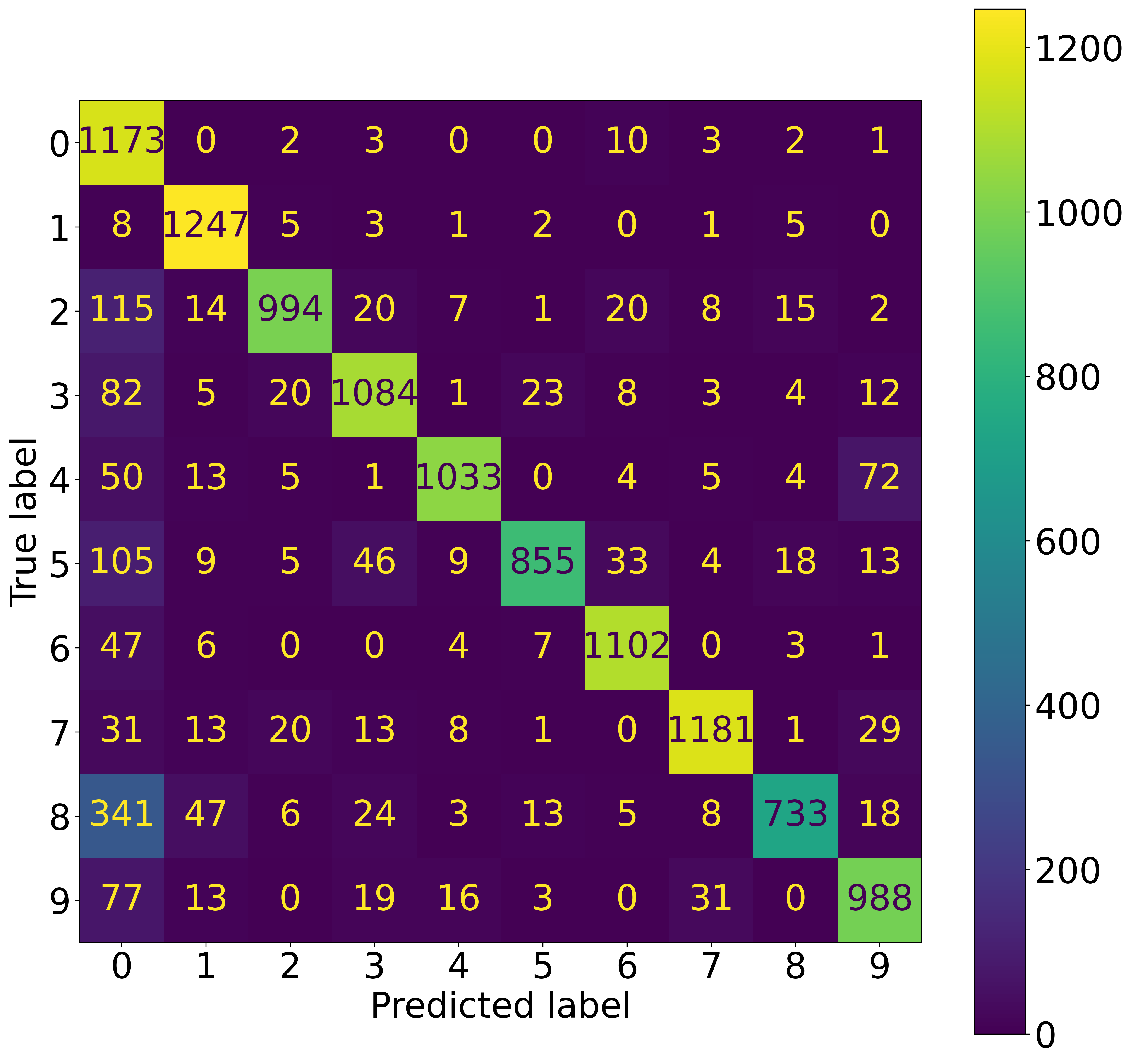}
\caption{Confusion matrix for perceptron, with hard sigmoid as activation function.}
\label{fig:cm2}
\end{minipage}
%\\
\qquad
\begin{minipage}{5cm}
\includegraphics[width=5cm]{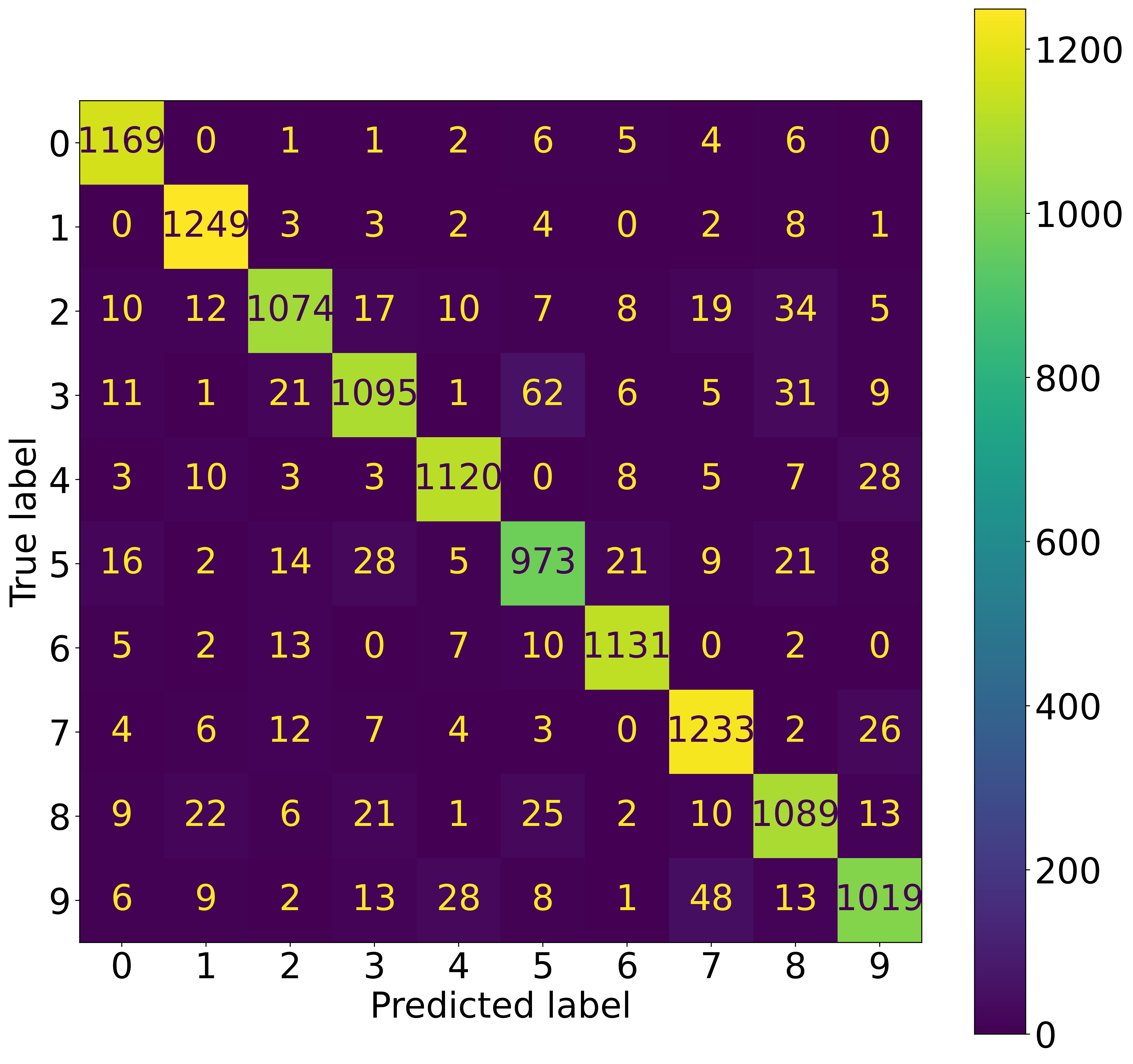}
\caption{Confusion matrix for perceptron, with softmax as activation function.}
\label{fig:cm4}
\end{minipage}

\end{figure}

We note that within the case of the perceptron based on TQFT some digits are predicted with lower-than-average precision, particularly 5 and 9. This behaviour can be explained because the probability amplitude in the semi-classical limit is sensitive to the topological characteristics of the examples. Within the case of the handwritten digits, a few pixels may separate circles and open chains, making classification less precise, as shown in Fig.~\ref{fig:cm1}-\ref{fig:cm4}.\\

\section{Perspective} \label{persp} 
\noindent 
The main contributions in dealing with the problem outlined by Refs.~\cite{zhang,zhang2017} can be subdivided into two classes: (i) theoretical approaches that try to understand the generalisation issue by proving a generalisation bound on the test error; (ii) phenomenological approaches motivated by experimentation --- as concerns a review of these approaches, see e.g. \cite{jiang2021methods}.

Further elaborating on the consequences of the results attained by \cite{zhang} a novel strategy has been developed in \cite{TQNN} and here that is rooted on the analogue framework provided by TQFT. This is an effective quantum theoretic approach that offers the pathway to address the problem of generalisation. The origin of this latter has been here related to the topological encoding of the input degrees of freedom by the network structure, achieved through pattern selection and parameter optimisation, in the semiclassical limit. The perspective we have then pushed forward here relies on the conjecture that generalisation happens as the analogue of the macroscopic manifestation of quantum effects.

A canonical experiment shedding light on our proposed outlook was carried out by Philipp Lenard, who unveiled the existence of the photoelectric effect. The occurrence of this effect, which only happens at frequencies of the impinging electromagnetic radiation (photons) that are above a certain threshold, provides a macroscopic manifestation of the existence of a quantised energy gap between electronic bounded and valence states, occurring in electric conductive materials.

This was indeed the framework adopted by Albert Einstein, who achieved a theoretical understanding of such a semiclassical effect grounding it on features of the newly developed theory of quanta. Analogously, generalisation can be addressed as the manifestation of a topological (quantum) encoding achieved by TQNNs. The texture of the webs of vertices and edges, which capture the topological structures of the graphs dual to the input data, implements the pattern \cite{TQNN}. For instance, the skeleton graph dual to the letter 'L' differs topologically from the one dual to the number '8', but metric properties alone distinguish among the number '8' and the symbol ‘$\infty$’. 

Generalisation is achieved in the TQNNs framework as a selection, induced by the quantum algorithm, of topological features that are the most adequate to the achievement of a specific task. These topological features are captured by the connectivity of graphs in the boundary states and of the vertices of the $2$-complexes. Thus in our proposed picture, the analog of the quantum states of the photoelectric effect --- the electrons that appear either in bounded or in valence energy levels, and the photons impinging the condenser’s plates of the conductor metal in the Lenard experiment --- are the quantum states represented by the cylindrical functionals of the boundary group elements (labelling the input data). These functionals are supported on the boundary graphs ($1$-complexes), and it is their functorial evolution that is captured by the TQN2C classifiers. Output boundary states represent instead the TQNN's ability to react to perturbations imposed by --- effectively, the queries posed by ---  the input training sets and test samples. The measure of success, for both training and test measures, is provided by the internal product among the boundary states, which is instantiated through the TQN2C functor, accounting for the evolution of the TQNN states \cite{TQNN}.

The novelty of our theoretical approach, in particular with respect to recent inspiring studies on the TQNN framework \cite{TQNN,fgm:22}, allows to consider a richer architecture that enables to associate machine learning concepts entailing complexity to the topological features the are coded therein. These properties include not only the inter-connectivity of the edges belonging to the graphs, but possibly also associated linking and knotting numbers, and the topological invariant properties of the 2-complexes spanned by the graphs’ evolution \cite{TQNN,fgm:22}. 

Within this framework, generalisation emerges from the optimisation of the topological structures, topological invariants (states’ sum) and quantum numbers (topological parameters), while the other parameters, which we may call metric, eventually instantiate effective macroscopic thresholds, such as in the photoelectric effect. Indeed, switching metric parameters on, does trigger the emergence of the topological features too --- see \cite{TQNN} for the details of the dual graphs selection out of the input data. Change of the graphs' topology is then achieved at the hidden layers by vertices structures implementing the TQNN evolution through the TQN2C. Furthermore, the volume of the input data set, increasing with the number of links and nodes, to which holonomies and intertwiner tensors respectively are associated, will play the analogue of the intensity of the radiation in the photoelectric effect (number of photons). While the dimensions of the spin-representations, assigned to the boundary TQNN states, namely the dimension of the Hilbert spaces associated to each link and node, will play an analogous role to the frequencies of the electromagnetic radiation. 

Through the definition of this architecture, TQNNs can capture the topological invariants from the training sets, which enables the identification of the correct output, once test samples are deployed. This happens through minimisation and optimisation of the classical action, namely by stationarising the path integral representation of the TQN2C classifiers. This procedure is reminiscent of the free energy principle \cite{friston:10, friston:13, friston:19, ffgl:22}: in the semiclassical limit the most important contributions to the path-integral evaluation of the classifier are the paths that are closest to the classical ones.

Asymptotically, in the semiclassical limit, a finite sample of labels suffices to the success of the generalisation process \cite{TQNN}. This observation suggests a different novel comprehension of the problem of generalisation. Indeed, we have also to remind that, along Zhang and others’ work, the generalisation process is independent on the regularisation of the dimensionality of the label sets that are involved. The proposed resolution of the problem may be then naturally achieved through the architectures of the TQNN states, as pointed out in \cite{TQNN}, while accounting for a class of TQNN states that are solely supported on graphs and $2$-complexes of reduced connectivity. 

\section{Conclusions} \label{concl}
\noindent
Moving from the innovative framework of \cite{TQNN}, we tackled the most relevant theoretical issue related to DNNs: how is it possible that DNNs are able to generalise and, therefore, learn? Understanding how generalisation works may allow to build a principled model of the operation of Deep Learning architectures. On the other hand, delving into the DNNs generalisation process from the perspective of topological quantum physics can provide the key to unprecedented technological implementations.

Considering first the heuristic case of one-node states, which can be treated in full analogy with standard quantum mechanics, and then extending the analysis to multi-node states of a quantum version of graph neural networks, i.e. topological quantum neural networks, we have shown that the origin of the problem of generalisation can be related to the topological encoding within the quantum graph neural network structure, achieved through path selection and parameter optimisation that corresponds to the semi-classical limit on quantum theories. 

We have hence provided an extended and intuitive explanations of the generalization process induced by TQNNs. Most importantly, we have presented a detailed numerical analysis of the perceptron case, as a proof of concept of how TQNN generalize, showing how it reproduces the results of DNNs, hence achieving generalisation. To better elucidate the generalisation power of TQNNs, we have provided a comparison between our results and results from standard DNN’s procedures. The semi-classical limit of TQNNs does not require any training using a direct computational approach comparable to the approach used by optimization algorithms (e.g. stochastic gradient descent) generally considered to be opaque. 

Thus, through our numerical analysis, we have shown that the generalization problem disappears, thus it is solved, if we adopt the TQNN/TQFT perspective. Technically and practically, there is no need for training, but one can rather calculate directly (not optimize) the weights of a trained quantum neural network, based on the probability amplitudes of TQFT. Still, one might argue that this is not computationally convenient in terms of accuracy, at least with our current technology, as we would need a quantum computer to achieve this result. Nonetheless, our approach provides an answer to the question of why the specific configuration obtained during training corresponds to a local minimum. The saddle point of the path integral is indeed a collection of points, and we find that training via stochastic gradient descent produces different minimal configurations because they correspond to different saddle point configurations. Notice however that, using quantum materials to implement our neural networks, high accuracy can be achieved (namely, no approximation errors, as in our numerical analyses) without training. Neural networks can be then designed, based on specific tasks, specifying configurations with the assignment of weights that correspond to those ones that appear in the path integral. There is no clearly optimization in such a procedure, and no unclear reasons of why a certain configuration is better than the other. 

Our proposed explanation may have a social and economic impact as well to the extent that it improves the trustworthiness of AI systems, and their practical and industrial applications. We further emphasize that the innovative flavour of our analysis comes from its own interdisciplinary features, which achieve constructing a bridge between traditional ML, with particular regard to DNNs, topological quantum physics and quantum field theory, and materials science.

  \section*{Acknowledgments}

We acknowledge Krid Jinklub for useful comments. AM acknowledges support by the National Science Foundation of China, through the grant No. 11875113, the Shanghai Municipality, through the grant No. KBH1512299, and by Fudan University, through the grant No. JJH1512105. ML acknowledges the support from National Science Foundation of China grant No.~12050410244.

% \bibliographystyle{IEEEtran}
% % \bibliographystyle{unsrtnat}
% % \bibliographystyle{model1-num-names}
% \bibliography{new_biblio.bib}
% %%\bibliography{main.bib}

%\vskip-3
%
%%\vskip -3\baselineskip plus -1fil

\appendix

\section{A dictionary for TQNNs}\label{dic}
\noindent 
Along the lines specified in \cite{TQNN}, we recall in this appendix the dictionary between TQNNs and the most relevant notions in standard machine learning, including DNN theory. This gives more context to the study conducted in this article. 

It is useful to restrict our focus to supervised learning. This latter task implements learning of a usually unknown function $f: D \rightarrow Y$ that maps an input set $D$ to an output set $Y$, and is based on a training set $S \subset D$ and a function $f^{\prime} : S \rightarrow Y$ that specifies example input-output pairs.  Considering $f: X \rightarrow Y$ as the (presumably random) function $r$ implemented by the network before the training, the learning algorithm can be specified 
%, as in Eq. \eqref{def-learn}, 
as an operation $\mathcal{L}: (r, f^{\prime}) \mapsto f$. A statistical learning framework for supervised learning can be then introduced, along the lines of \cite{shalev-shwartz}, as well as some standard definitions for DNN that we list below. 
\begin{itemize}
\item
Sample complexity:\\
It represents the number of training-samples (i.e. $Card(S)$) that a learning algorithm needs in order to learn successfully a family of target functions.

\item
Model capacity:\\
It is the ability of the model to fit a wide variety of functions; in particular, it specifies the class of functions $\mathcal{H}$ (the hypothesis class) from which the learning algorithm $\mathcal{L}$ can choose the specific function {${h}$}.

\item
Overfitting:\\
A model is overfitting when the gap between training error and test error is too large; this phenomenon occurs when the model learns the training function $f^{\prime}$ but $\mathcal{L}$ incorrectly maps $(r, f^{\prime}) \mapsto h \neq f$, i.e. the trained network generalises to the wrong function $h$ and fails to predict future observations (i.e. additional samples from $D$) reliably.  The training function $f^{\prime}$ has been merely ``memorised'' to the extent that $h$ is incorrect (e.g. random) on $D$ outside of the training sample $S$.

\item
Underfitting:\\
A model is underfitting when it is not able to achieve a sufficiently low error on the training function $f^{\prime}$; this phenomenon occurs when the model does not adequately capture the underlying structure of the training data set and, therefore, may also fail to predict future observations reliably. 

\item
Bias:\\
It is the restriction of the learning system towards choosing a classifier or predictor {${h}$} from a specific class of functions $\mathcal{H}$ (the hypothesis class).

\item
Empirical Risk Minimisation (ERM):\\
It consists in minimising the error on the set of training data (the ``empirical'' risk), with the hope that the training data is enough representative of the real distribution (the ``true'' risk).

\item
Generalisation:\\
It is conceived as the ability of the learner to find a predictor, i.e. an embedding $S \rightarrow D$, which is able to enlarge successfully its own predictions from the training samples to the test or unseen samples.

\end{itemize}

These notions can be reformulated into the dictionary of TQNNs.

\begin{itemize}

\item 
Sample complexity:\\
It is a measure of the Hilbert-space of the entire spin-network state that is supported on a specific graph $\Gamma$. It is then dependent on the connectivity of the graph (nodes and links of each graph, i.e. the multiplicity of connectivity that characterizes the graph $\Gamma$) and on the dimensionality of the Hilbert spaces connected to each link and node. In this sense complexity, once extended to the different classes of graphs corresponding to the training set, provides a measure of the entropy of the set. Therefore, in the TQNN framework, the notion of “complexity” has a wider meaning than its counterpart in DNN, for which the sample complexity is nothing but the size of the training set. This is summarised in the expression for the dimension of the Hilbert space $\mathcal{H}_{\Gamma}$ of the (whole) spin-network supported on $\Gamma$, namely 
$$
{\rm dim}[\mathcal{H}_{\Gamma}]=\oplus_{j_l} \otimes_n \otimes_{l\in \partial n} \, {\rm dim}[\mathcal{H}_{j_l}].
$$
This directly encodes both the size of the maximal graph where the input/output states live, as well as the algebro/analytical structure used in the TQFT from which the corresponding TQNN arises, as encoded by the dimensionality of the Hilbert spaces $\mathcal H_j$, for instance;

\item
Model capacity:\\
It is now distinguished into topological model capacity and metric model capacity, the latter being the extension of the definition provided for DNNs to the context of TQNNs.\\
i) Topological model capacity: It is quantified in terms of the interconnectivity of the graph $\Gamma$. It depends on the topological structure of the graphic support $\Gamma$ of the spin-network states, and neither on the dimensionality of the Hilbert space of the irreducible representations nor on the intertwiner quantum numbers, respectively assigned to each link and node of $\Gamma$; in other words, it depends on the total valence $V$ of $\Gamma$, defined in terms of the valences $v_n$ of each node of $\Gamma$ through the expression
$$
V=\sum_n v_n \, ;
$$
ii) Metric model capacity: At fixed graph $\Gamma$, it depends on the dimensionality of the Hilbert space of the irreducible representations and intertwiner quantum numbers assigned to $\Gamma$;\\
iii) Combined model capacity: It combines the topological and metric capacity, so as to mimic the standard DNN notion of model capacity. It provides a a representation of the double Belkin curve as in Figure~\ref{DC}.

\item
Overfitting:\\
As pointed out in Section~\ref{TQNN}, in the semi-classical limit, the integrals that allow us to compute the transition amplitudes that characterise a TQFT are interpreted as a ``sum over all the geometries'' of the ground topological manifold, where the integrand is some approximation of the Einstein-Hilbert action. During the learning process, then a TQNN learns how to select certain geometries with respect to certain others in order to maximise certain transition amplitudes corresponding to ``a more suitable'' classification. The information available to make this selection during the learning process is that given by the metric data, namely the irreducible representations and intertwiner quantum numbers assigned to the TQNNs graphs, and by the connectivity of the input graphs/spin-networks and their given correlation $f^{\prime}$ with the label set $Y$. Keeping metric data fixed, if $f^{\prime}$ is insufficiently representative of the target function $f$, the TQNN may only partially capture the topological structure of the full input set $D$ and therefore be unlikely to classify correctly spin-network states that are not part of, or are significantly dissimilar from those contained in, the training set $S$. Conversely, when connectivity is kept fixed, overfitting follows the standard behaviour of DNNs. Then a classical U-shaped risk curve describes
the trade-off between underfitting and overfitting;

\item
Underfitting:\\
It represents the converse of the overfitting scenario. The geometries that have been selected in the learning process do not correspond to the graphs $\Gamma$ at the starting point. Less information channels (links) are present, and lower dimensionality of the information channels (dimensions of the Hilbert space associated to each holonomy) as well. As a consequence, the TQNN cannot fit the training set and may therefore also fail to predict future observations reliably;
 
 \item
Bias:\\
It amounts to the predisposition of the spin-network to account for a specific set of data; it depends on the topological structure of the spin-network states, encoded in the connectivity properties of input $\Gamma$'s and on the specific realisation of the TQNN quantum state, i.e. on the weight of the quantum state on the spin-networks basis elements of the Hilbert space.

\item
Empirical Risk Minimisation (ERM):\\
It is the variance of the Gaussian distribution of the irreducible representations assigned to the holonomies on the links in the semi-classical limit, i.e.
$$
{\rm ERM}:= \sum_l \frac{(j_l -\bar{j}_l)^2}{2 L}
\,,
$$
with $L$ equal to the total number of links.

\item
Generalisation:\\
It is the behavior of the system in response to test or unseen data analogous to a functor (amplitude) either from a boundary spin-network to another boundary spin-network, or from a boundary spin-network to a complex number.
This is determined by the geometries that have been selected as the most representative of a certain training sample during the learning process. This is in practice captured by the parameters that give higher relevance, in the integral computing the transition amplitudes in a TQNN, to certain boundary transitions, while suppress others. These parameters are determined by (i) connectivity of 1- and 2-complexes (nodes and links, vertices and edges respectively), (ii) linking and knotting (e.g. for loops in a different Hilbert space representation), and (iii) states’ sum (as a global topological charge, invariant under refinement of the triangulation, i.e. invariant under refinement of the data/group elements/intertwiners assigned to the links and the nodes). In \cite{TQNN}, parameters enter the expression for the amplitudes thanks to the coherent states formalism. For $U_l$ elements of a group $G$ we may resort to the formula for the partition function of the model:

\begin{eqnarray} \label{funbis1}
\mathcal{Z}_\mathcal{C}(U_l)= \int_{{\rm SU}(2)^{2(E-L)-V} } dU_{v(e)} \, \int_{{\rm SU}(2)^{\mathcal{V}-L}} dU_f\, \nonumber\\ 
\times \prod_f\, \mathcal{K}_{f*}(U_{e*},U_f) \,, 
\end{eqnarray}
where the ``face amplitude'' casts 
\begin{eqnarray} \label{funbis3}
\mathcal{K}_{f*}(U_{e*},U_f)\equiv 
\sum_{j_{f*}} \,  \Delta_{j_{f*}} \,  \chi^{\scriptscriptstyle j_{\!f*}}\!\Big(\!\prod_{e*\in\partial f}U_{e*}\!\Big) \, \nonumber\\ 
\times \prod_{e*\in\partial f}\!\chi^{\scriptscriptstyle j_{\!f*}}(U_f)\,.
\end{eqnarray}

\end{itemize}

 Finally, from the definitions of the present article, we can provide the meaning of Learner's input and output in the context of TQNN.

\begin{itemize}
\item
Learner’s input:\\
i) The domain set $D$: It corresponds to links $l$ and nodes $n$, and attached holonomies $U_l$ and invariant tensors $\iota_n$ respectively along the links and at the nodes: it is concisely denoted as a state of the Hilbert space of the theory:   
%\begin{equation}
$$
\Psi_{\Gamma ; \{j_l\}, \{\iota_n\}}[A] \equiv \Psi_\Gamma(U_l, \iota_n) := | \Gamma; \{j_l\}, \{\iota_n\}  \rangle ;    
$$
%\end{equation}
ii) The label set $Y$: It is a set of topological charges and quantum numbers, with which the 2-complex is endowed; for instance, recalling the group-isomorphism $\pi_3(S_3)$, for the mapping individuated by the homotopy group  $\pi_3(S_3)=\mathbb{Z}$ the winding number $w$ is defined as the integral over the SU$(2)$ group element 
$$
w=\frac{1}{24 \pi^2} \int_{\rm SU(2)} dU ;    
$$
\\
iii) The training data $S$: it is the union of the (initial) boundary colored graphs together with the topological invariants associated to them through the TQN2C functorial action.

\item
Learner’s output:\\
It is a prediction rule, i.e. the TQN2C that identifies the topological charges of the boundary states (training/test samples) and thus implements the classifier; for $\gamma$ a $1$-complex supporting a disjoint boundary state, and $\mathcal{C}$ a $2$-complex with boundaries $\partial \mathcal{C}=\gamma$, the classifier is captured by the probability amplitude that results from the internal product 
$$
 \mathcal{A}=\langle  \gamma; \{j_l\}, \{\iota_n\} | \, | \mathcal{Z}_{\mathcal{C}, \partial \mathcal{C}=\gamma } ;  \{j_l\}, \{\iota_n\}\rangle \,,  
$$
or, once the physical projector $\mathcal{P}$ has been determined, $\mathcal{A}=\langle \psi_{\rm out}| \mathcal{P} \psi_{\rm in} \rangle$, for generic boundary states $\psi_{\rm in}$ and $\psi_{\rm out}$.

\end{itemize}

These definitions illustrate, in a very explicit way, the difference between how a TQNN ``sees'' the input and label sets --- and hence the semantics that it assigns to these sets --- and how we, as human engineers, see them.  From an explainable AI (xAI) perspective, the semantics assigned by the TQNN is effectively uninterpretable.  Hence as noted above, what our current approach provides is not an explanation of how generalisation has worked in any particular case, but rather an assurance, up to relevant conditions, that it has worked.

\bibliographystyle{IEEEtran}%[0]
\bibliography{IEEEabrv,new_biblio.bib}
% \endgroup

\end{document}